\documentclass[11pt]{article}
\usepackage[pagewise]{lineno}

\title{Parallel Token Swapping  for Qubit Routing}

\usepackage{amsmath,amsfonts,amssymb,url,graphicx,subcaption,float,booktabs,bbm,enumitem,amsthm}
\usepackage{amsmath}
\usepackage{amsfonts}
\usepackage{amssymb}
\usepackage[T1]{fontenc}
\usepackage{lmodern}
\usepackage{xcolor}
\usepackage{algorithm}
\usepackage{algpseudocodex}
\usepackage{tabularx,environ}
\usepackage[]{mdframed}
\usepackage{authblk}
\usepackage{subcaption}

\newtheorem{thm}{Theorem}
\newtheorem{lemma}[thm]{Lemma}

\newtheorem{definition}[thm]{Definition}
\newtheorem{observation}[thm]{Observation}
\newtheorem{remark}[thm]{Remark}

\newenvironment{claimproof}{\begin{proof}}{$\Box$\end{proof}}
\newtheorem{claim}{Claim}
\newtheorem{oldthm}{Theorem}

\makeatletter
\newcommand{\problemtitle}[1]{\gdef\@problemtitle{#1}}
\newcommand{\probleminput}[1]{\gdef\@probleminput{#1}}
\newcommand{\problemquestion}[1]{\gdef\@problemquestion{#1}}
\NewEnviron{problem}{
  \problemtitle{}\probleminput{}\problemquestion{}
  \BODY
  \par\addvspace{.5\baselineskip}
  \noindent
  \begin{tabularx}{\textwidth}{@{\hspace{\parindent}} l X c}
    \multicolumn{2}{@{\hspace{\parindent}}l}{\@problemtitle} \\
    \textbf{Input:} & \@probleminput \\
    \textbf{Output:} & \@problemquestion
  \end{tabularx}
  \par\addvspace{.5\baselineskip}
}
\makeatother

\newcommand{\matching}{\ensuremath{M}}
\newcommand{\config}{\ensuremath{C}}
\newcommand{\tokenset}{\ensuremath{T}}

\newcommand{\Id}{\ensuremath{Id}}

\newcommand{\PTSx}{({\fontfamily{cmss}\selectfont PTS})}
\newcommand{\PTS}{\PTSx~}

\newcommand{\Pot}{\ensuremath{\Phi}}
\newcommand{\new}[1]{{\textcolor{black}{#1}}}

\begin{document}
\author[1]{Ishan Bansal}
\author[2]{Oktay G\"unl\"uk}
\author[3]{Richard Shapley}
\affil[1]{Amazon, Bellevue, WA\footnote{This work was completed prior to being affiliated with Amazon and does not relate to the current role at Amazon.}}

\affil[2]{Industrial and Systems Engineering, Georgia Institute of Technology}
\affil[3]{Operations Research and Information Engineering, Cornell University}%
\maketitle              
\begin{abstract}
In this paper we study a combinatorial reconfiguration problem that involves finding an optimal sequence of swaps to move an initial configuration of tokens that are placed on the vertices of a graph to a final desired one. This problem arises as a crucial step in reducing the depth of a quantum circuit when compiling a quantum algorithm. We provide the first known constant factor approximation algorithms for the parallel token swapping problem on graph topologies that are commonly found in modern quantum computers, including cycle graphs, subdivided star graphs, and grid graphs. We also study the so-called stretch factor of a natural lower bound to the problem, which has been shown to be useful when designing heuristics for the qubit routing problem. Finally, we study the colored version of this reconfiguration problem where some tokens share the same color and are considered indistinguishable.

\end{abstract}
%
%
%


\newcount\mycount

\section{Introduction}





Combinatorial reconfiguration problems have been studied for decades due to their applications in puzzle solving, robotics and motion planning, string rearrangement, optimal qubit routing, and troop and swarm reassignments. The general paradigm is the following: we are given an initial configuration of pieces/items/agents, along with a final desired configuration, and we wish to find an efficient sequence of permissible moves that modify the initial configuration to obtain the desired one. One model used to mathematically analyze such combinatorial reconfiguration problems is called token reconfiguration. In this model, we are given an undirected connected graph $G$, with physical tokens placed on its vertices in some initial configuration. We wish to route these tokens to a final desired configuration using a shortest sequence of permissible token swaps. Different definitions of permissible token swaps yield many interesting variants of the token reconfiguration problem. In the token swapping problem introduced by Yamanaka et al. \cite{yamanaka2015}, swaps are permissible only if the tokens are on vertices connected by an edge of the graph $G$; the objective in this setting is to find the minimum number of swaps required to obtain the final configuration from the initial one.

Motivated by recent applications in qubit routing on quantum computers, we study the parallel token swapping problem first introduced by Alon et al. \cite{alon1994}. Here, one is permitted to swap multiple pairs of pairwise disjoint adjacent tokens at the same time. The parallel token swapping problem has been shown to \new{be} NP-\textsc{Hard} even on subdivided stars \cite{aichholzer2022}; we provide the first known constant factor approximations for several common graph topologies arising as substructures in modern quantum computers \cite{jurcevic2021demonstration,nannicini2022optimal,wagner2023improving,cowtan2019qubit}. In addition, we compare our approximation factors to the so called \textit{stretch factor} of a natural lower bound (the maximum geodesic distance of a token between its positions in the initial and final configurations) and show that our approximation factors are tight with respect to this lower bound in some cases. The stretch factor refers to the maximum ratio of $OPT/LB$ over all instances where $OPT$ is an optimal solution and $LB$ is the lower bound.

\subsection{Motivation}\label{sec:motivation}

Quantum algorithms are usually designed under the assumption that any two qubits in a quantum computer can interact via a quantum gate. However, in practical \new{hardware}, physical qubits are laid out in a two-dimensional or three-dimensional topology and only adjacent qubits can interact via gates. Fortunately, this discrepancy can be handled by using SWAP gates, which switch the positions of two adjacent qubits (we refer the reader to \cite{nielsen2010quantum} for a detailed discussion). This process is referred to as qubit routing, and is a necessary step for \new{many quantum compilers}. SWAP gates should be added carefully, as they increase the size and depth of the quantum algorithm, which adversely affects the fidelity of a quantum circuit (see, for example \cite{johnstun2021understanding,jurcevic2021demonstration}). Hence, the goal of the qubit routing problem is to find an initial positioning of the qubits and a sequence of SWAP gates that minimizes the size and/or depth of the quantum circuit. The initial allocation subproblem is equivalent to the graph isomorphism problem \cite{siraichi2018qubit,childs2019circuit} while the problem of swapping qubits is equivalent to the the token swapping problem \cite{yamanaka2015}. The parallel token swapping problem optimizes the depth of the quantum circuit.

The qubit routing problem has proven to be extremely hard, owing in part to the complexity of the subgraph isomorphism and token swapping problems. Exact algorithms can be found in the works of \cite{siraichi2018qubit,nannicini2022optimal,wille2019mapping}. These algorithms, however, have highly exponential running times that do not scale well with the size of the quantum computer. Consequently, there is a large body of research on developing empirically good heuristics for the problem \cite{cowtan2019qubit,sivarajah2020t,zulehner2018efficient,li2019tackling,qiskit2019open,kissinger2019cnot,moro2021quantum,pozzi2022using,sinha2022qubit,molavi2022qubit,wagner2023improving}. We refer the reader to the following survey on algorithms for the qubit routing problem \cite{barnes2023survey}. Very recently, Wagner at al. \cite{wagner2023improving} considered the problem of optimizing the size of the quantum algorithm and observed that approximation algorithms for the token swapping problem can be leveraged to provide superior heuristics for the qubit routing problem. In particular, the previously known 4-approximation algorithm for the token swapping problem \cite{miltzow2016} can be used to guide the heuristic when deciding on qubit allocations. 
They attempted to extend their results to the problem of optimizing the depth of the quantum circuit, but were limited by the lack of good approximation algorithms for the parallel token swapping problem. Their methodology is to use the 4-approximation algorithm for the non-parallel token swapping problem and combine gates into layers sequentially. However, optimal solutions to the parallel token swapping problem can be very different from optimal solutions to the token swapping problem, bringing to light the importance of designing good approximation algorithms for the parallel token swapping problem. In this work, we provide approximation algorithms for the parallel token swapping problem on common graphs arising as substructures in current quantum topologies (\cite{jurcevic2021demonstration,nannicini2022optimal,wagner2023improving,cowtan2019qubit}). For example IBM's heavy hex lattice quantum architecture can be viewed as a union of cycle graphs and also as a union of subdivided star graphs. \new{We note that designing good algorithms for more complex graph topologies (including IBM's heavy hex and Google's diagonal grid) using algorithms for its subgraphs provided in this paper requires further research.}

A complete solution to the qubit routing problem requires a sequence of qubit allocations under which the quantum gates can be implemented, and also a sequence of swaps to move from one allocation to another. The stretch factor of lower bounds for the token swapping problem is a crucial tool linking the two parts. If the stretch factor is reasonably bounded, one can use this lower bound as a proxy to determine good qubit allocations. For instance, Wagner et al. \cite{wagner2023improving} designed a heuristic for the qubit routing problem using the half sum lower bound of Miltzow et al. \cite{miltzow2016}, which was shown to have a stretch factor of 4 for the token swapping problem. They were able to use this lower bound as an objective in the integer program of \cite{nannicini2022optimal} to develop superior heuristics for the qubit routing problem. Hence, a careful analysis of the \new{stretch} factor of natural linear lower bounds is essential for a better understanding of the qubit routing problem. In the parallel token swapping problem, the natural linear lower bound is the maximum geodesic distance of a token between its initial and final configurations. We study the stretch factor of this lower bound on the graph topologies mentioned earlier.

\new{We should note that qubit routing is also possible without SWAP gates. In particular, there are Hamiltonian routing methods that are stronger than SWAP based routing. There are also methods based on local operations, measurement and feedback called LOCC that are akin to qubit teleportation. We refer the reader to \cite{PRXQuantum.4.010313} and \cite{PhysRevResearch.6.033313} for discussions on the advantages and disadvantages of these methods. These other routing methods are not considered in this paper.}

\subsection{Our Contributions}\label{subsec:Contributions}


We begin with a result showing that the stretch factor of the maximum geodesic distance lower bound can be arbitrarily bad for the parallel token swapping problem \PTSx. Let $d_i$ be the geodesic distance of the $i^{th}$ token between its initial and final positions and let $n$ be the number of vertices (or tokens) in the graph. Let $d_{max} = \max\{d_1,\ldots,d_n\}$. We show that any function of these distances, $f(d_1,\ldots,d_n)$ has an arbitrarily bad stretch factor (maximum OPT/$f(d_1,\ldots,d_n)$ ratio over all instances).  This is in stark contrast to the non-parallel token swapping problem where the lower bound $(d_1+\cdots+d_n)/2$ has a stretch factor of 4 on all graphs, which was leveraged by \cite{miltzow2016} to provide a 4-approximation algorithm.

\begin{thm}\label{thm:hardness}
    Let $f(d_1,\ldots,d_n)$ be a valid lower bound for \PTS for some function $f$. Then the stretch factor of this lower bound is $\Omega(n)$ for graphs on $n$ vertices.
\end{thm}

We prove this by providing two instances of the problem with the same distance vector $(d_1,\ldots,d_n)$. The first instance is on the complete graph and it is well known that the optimal solution on the complete graph uses at most two steps. The second instance is on the cycle graph and we argue that any solution must use at least $n-1$ steps. \new{The above result suggests that the design of good approximation algorithms for \PTS must exploit the topology of the underlying graph.} Due to this, we focus our attention to specific graph topologies arising in quantum computers, beginning with a result on cycle graphs. Throughout, we use $OPT$ to denote  the optimal solution value of a given instance of \PTSx.

\begin{thm}\label{thm:cycle}
    Let $G$ be an $n$-cycle, then,
    \begin{enumerate}
        \item[(a)] If $n$ is even, we can find a solution with value at most $2 OPT$ in polynomial time.
        \item[(b)] If $n$ is odd, we can find a solution with value at most $2 OPT +1$ in polynomial time.
    \end{enumerate}
\end{thm}

The proof of this result involves a reworking of the approximation algorithm on line graphs by \cite{kawahara2017}. We repeatedly `cut' the cycle graph into a line graph and apply a modified version of the algorithm in \cite{kawahara2017} to argue that some swaps are `reasonable' and can be assumed to occur in an optimal solution. However, we observe that this does not always work since tokens on a cycle can loop around. In such a case, we show that the optimal solution must be sufficiently large. We can then design an approximation algorithm that runs two different algorithms and outputs the better solution of the two.

Next, we focus our attention on sub-divided star graphs. These are graphs where exactly one vertex has degree greater than 2. Hence, the graph is a collection of line subgraphs all connected at a single vertex.

\begin{thm}\label{thm:star}
    Let $G$ be a sub-divided star with maximum degree $h$, then
    we can find a solution with value at most $4OPT + \min\{OPT,h\} + 1$ in polynomial time.
\end{thm}

The algorithm here works in three phases. In the first two phases, we move tokens around so that each token is in the line subgraph where its final position belongs. We then reconfigure each line subgraph in parallel. We show that the first two phases can be done near optimally since the central vertex with high degree acts as a bottleneck for any solution including the optimal one. To bound the number of steps required in the third phase, we modify any optimal solution so that the configuration at the end of the second phase of our algorithm and the configuration at the end of a similar phase in the modified optimal solution are not too different. Hence, we can reconfigure the algorithm's configuration to an optimal one without losing too many steps. 

Next, we look at grid graphs. Here we design two different algorithms that work better depending on the size of the grid graph.

\begin{thm}\label{thm:grid}
    Let $G$ be a $h \times n$ grid graph with $h \leq n$, then
    \begin{enumerate}
        \item[(a)] If $h = 2$, we can find a solution with value \new{at most $2OPT + 2$ in} polynomial time.
        \item[(b)] If $h \geq 3$, we can find a solution with value at most $2 OPT + 2h$ in polynomial time. 
    \end{enumerate}
\end{thm}

\begin{table}[t]
    \centering
    \begin{tabular}{|c|c|c|c|}
        \hline
        ~ &  \PTS & stretch & colored \PTS\\
        \hline
        Line Graphs & $OPT + 1$ \cite{kawahara2017} & $\mathbf{2}$ & $OPT + 1$ \cite{kawahara2017}\\
        \hline
        Even Cycle Graphs & $\mathbf{2OPT}$ & $\mathbf{n}$ & $\mathbf{2OPT}$ {\small incomplete \PTS}\\
        \hline
        Odd Cycle Graphs & $\mathbf{2OPT +1}$ & $\mathbf{n}$ & $\mathbf{2OPT+1}$ {\small incomplete \PTS}\\
        \hline
        Subdivided Stars Graphs & $\mathbf{4OPT + \min\{OPT,h\} + 1}$ & $\mathbf{\Theta(h)}$ & $\mathbf{4OPT + \min\{OPT,h\}+ 1} $\\
        \hline
        Ladder Graphs & $\mathbf{2OPT + 2}$ & --- & $\mathbf{2OPT + 2}$ \\
        \hline
        Grid Graphs & $\mathbf{2OPT + 2h}$ & --- & $\mathbf{2OPT + 2h}$ \\
        \hline
    \end{tabular}
    \caption{A summary of the current state of the art. The second and fourth columns list the best known approximation algorithm for \PTS and colored \PTS respectively, and the third column lists the stretch factor of the $d_{max}$ lower bound. A boldface indicates work in this paper. Previous best factors were $O(|V|)$ \cite{alon1994,zhang1999}.}
    \label{table:results}
\end{table}

The algorithm when $h = 2$ works by isolating a spanning path from the grid graph and utilizing the algorithm in \cite{kawahara2017} to reconfigure the tokens along this line subgraph. The optimal solution however, may utilize the edges in the grid graph that are not on this path. We show that \new{those swaps can be simulated by} swaps on the path without spending too many additional steps. This algorithm does not scale well with $h$. Instead, for larger values of $h$, we perform a three phase algorithm where we reconfigure token along rows, then column, and then rows again to obtain the final configuration. At the end of the first phase, we ensure that each column has exactly one token whose position in the final configuration is in row $i$, for any $i$. We then reconfigure the columns to bring each token to its correct row, and finally reconfigure the rows in parallel. The phases along rows can be bounded by the size of the rows ($h$), while the phase along columns can be bounded by our next result on the stretch factor of line graphs.

We emphasize that to the best of our knowledge, the best known approximation factors prior to our work for all the cases we consider was $O(|V|)$ \cite{alon1994,zhang1999}.

\begin{thm}\label{thm:stretch}
    
    \begin{enumerate}
        \item[(a)] The stretch factor of $d_{max}$ for line graphs is 2.
        \item[(b)] The stretch factor of $d_{max}$ for $n$-cycle graphs is between $n$ and $n-1$.
        \item[(c)] The stretch factor of $d_{max}$ for subdivided star graphs is $\Theta(h)$ where $h$ is the maximum degree of a vertex in the graph.
    \end{enumerate}
\end{thm}

The stretch factor for line graphs is obtained by designing a potential function on the tokens in any configuration, that evaluates to zero in the final configuration. We show that this potential can be as large as $2d_{max}-1$ and can be reduced by one in every iteration (except the first one). We also show an instance where the optimal solution makes at least $2d_{max}-2$ steps and hence, the stretch factor on line graphs is 2 asymptotically. For cycle graphs, our instance in the proof of Theorem~\ref{thm:hardness} immediately provides the result. For subdivided stars, we show that our algorithm in the proof of Theorem~\ref{thm:star} proves that the stretch factor is $O(h)$. A simple instance on $n$ vertices without any 2-degree vertices shows that the stretch factor is $\Omega(h)$.

Next, we extend our results to the colored versions of token swapping problem. The first one is the colored \PTS, where each token and each vertex is assigned a color, and every token should be routed to a vertex that shares its color. In this setting the number of colors can be strictly less than the number of vertices. A \PTS problem on a graph with $n$ vertices can be viewed as a colored \PTS on the same graph with $n$ colors where each token has a unique color.

Finally, we consider a variant of \PTS when the number of tokens is smaller than the number of nodes in the graph. This happens, for instance, if some qubits are not operational on the quantum computer. In this case, all tokens can be assigned a unique color and \new{all} empty positions can be assigned a common color and dummy tokens with that  color can be added in. We refer to this special case of the colored \PTS as the incomplete \PTSx.


\begin{thm}\label{thm:color}
    \begin{enumerate}
        \item[(a)] If $G$ is a cycle, we can find a solution with value at most $2 OPT + 1$ for the incomplete \PTS in polynomial time.
        \item[(b)] If $G$ is a sub-divided star, we can find a solution with value at most $4OPT + \min\{OPT,h\} + 1$ for the colored \PTS in polynomial time.
        \item[(c)] If $G$ is an $h \times n$ grid graph with $h \leq n$, we can find a solution with value at most $2 OPT + 2h$ for the colored \PTS in polynomial time.
    \end{enumerate}
\end{thm}

These results almost immediately follow from our previous theorems. However, a crucial first step is to assign a concrete final position for each token, thus converting the problem to an instance of the uncolored \PTSx. For the cycle graph, this is done by `guessing' the final configuration. We show that there are at most $n$ good guesses and so we can enumerate over all of them. For subdivided stars, we show that any configuration that does not require tokens of the same color to cross each other works. Hence, a greedy approach can be deployed. For the grid graphs, we solve a perfect matching problem to find the final configuration. Table~\ref{table:results} gives a summary of the current best known results. 

\subsection{Related Work}

In the non-parallel token swapping setting, Yamanaka et al. \cite{yamanaka2015} showed that not only does some sequence of swaps exist to rearrange tokens between any two configurations, but the number of swaps required can be bounded by $O(n^2)$. Miltzow et al. \cite{miltzow2016} proved that the token swapping problem is both NP-\textsc{Complete} and APX-\textsc{Hard}, but provide a 2-approximation for the problem when the underlying graph $G$ is a tree, which they extend to a 4-approximation for arbitrary graphs $G$. In the colored setting, Miltzow et al. \cite{miltzow2016} were able to achieve the same approximation factor as in the uncolored setting. Bonnet et al. \cite{bonnet2018complexity} showed that the problem is also $W[1]$-\textsc{Hard} and remains NP-\textsc{Hard} on graphs with constant treewidth. They also showed that the problem is fixed parameter tractable on nowhere dense graphs. Polynomial time exact algorithms are known for specific graphs like cliques \cite{cayley1849lxxvii}, line graphs \cite{knut1973art}, cycles \cite{jerrum1985complexity}, star graphs \cite{portier1990whitney}, brooms \cite{vaughan1999factoring}, complete bipartite graphs \cite{yamanaka2015}, and complete split graphs \cite{yasui2015swapping}.

The parallel token swapping problem was introduced by Alon et al. \cite{alon1994} and they showed that any instance on an $n$-vertex graph can be solved in $3n$ steps. This was later improved to $3n/2 + O(\log n)$ by Zhang \cite{zhang1999}. Kawahara et al. \cite{kawahara2017} showed that it is NP-\textsc{Hard} to decide if the optimal solution value is equal to three but it is decidable in polynomial time if the optimal solution value is equal to two. They also showed that on a line graph, there exists an approximation algorithm that outputs a solution with value $OPT + 1$.  Aicholzer et al. \cite{aichholzer2022} showed that the problem is NP-\textsc{Hard} even on subdivided stars. Not much else is known about the problem. Even the basic question ``Is the problem NP-\textsc{Hard} on line graphs?'' has been open for over three decades. We should point out that the introduction of \cite{aichholzer2022} claims that \cite{demaine2019coordinated} provided an $O(1)$-approximation algorithm for the problem on grid graphs. However, this claim is incorrect since \cite{demaine2019coordinated} considers a variant of the token reconfiguration problem where one is allowed the rotate the tokens along a cycle in one step. This is crucially used in steps 3 and 4 of their algorithm.

\section{Preliminaries}\label{sec:prelim}

Given a graph $G = (V,E)$, we define a set of \textbf{tokens} on $V$ to be a duplication of the vertex set denoted $\tokenset$. A \textbf{configuration} $\config$ is a bijection $\config: V \to \tokenset$. This bijection can be interpreted as a placement of the tokens on the vertices. In particular, $\config(v)$ refers to the token that lies on vertex $v$, and $\config^{-1}(t)$ refers to the vertex where token $t$ lies. As $\tokenset$ is a copy of the vertex set, without loss of generality, we use the canonical identity bijection $\Id$ to be the final configuration.
Let $M\subseteq E$ be a matching.
We say a matching $\matching$ \textbf{routes} a configuration $\config$ to obtain a new configuration denoted $\matching\config$ defined by 
\[\matching\config(v) = \begin{cases} \config(w), & \text{if } (v,w) \in \matching \\
\config(v), & \text{\new{otherwise.}}\end{cases}\] 
Similarly, a sequence of matchings $(\matching_1,\ldots,\matching_k)$ routes a configuration $\config$ to obtain the new configuration $\matching_k\cdots\matching_1\config$. With this notation, the parallel token swapping can be formally described as, \\

\begin{problem}
    \problemtitle{\large \textit{Parallel Token Swapping Problem \PTS}}
    \probleminput{A graph $G = (V,E)$ and an initial configuration $\config$.}
    \problemquestion{A sequence of matchings $(\matching_1,\ldots,\matching_k)$ with minimum size ($k$) such that $\matching_k\cdots\matching_1\config = \Id$.}
\end{problem}

\new{\begin{remark}\label{rem:consecutive_matchings_nonintersecting}
    Note that for the purposes of minimizing the size of the sequence of matchings $k$, we can assume that no pair of consecutive matchings $(M_i,M_{i+1})$ shares a common edge $e=(u,v)$. Indeed if $e\in M_i \cap M_{i+1}$, then $M_{i+1}M_iC' = M'_{i+1} M'_{i} C'$ for any configuration $C'$ where $M'_{i+1} = M_{i+1}\setminus\{e\}$ and $M'_{i} = M_{i}\setminus\{e\}$. Hence, without loss of generality, we shall assume throughout this paper that for any sequence of matchings $M_i \cap M_{i+1} = \emptyset$.
\end{remark}}

A useful measure of a good lower bound for an optimization problem is called the stretch factor. This measures the worst-case ratio of the lower bound and an optimal solution, as formally defined below.

\begin{definition}[Lower Bound]
    A function $\ell$ that takes as input an instance of \PTS described by $(G,C)$ and outputs a real number is called a lower bound to the \PTS problem if $$OPT(G,C) \geq \ell(G,C)$$ for all instances $(G,C)$ where $OPT$ is the optimal solution size.
\end{definition}

\begin{definition}[Stretch Factor]\label{def:stretch}
    Let $\ell$ be a lower bound to the \PTS problem. The stretch factor of the lower bound $\ell$ is $$\max_{(G,\config)} \frac{OPT(G,\config)}{\ell(G,\config)}.$$
\end{definition}
For a given configuration $\config$ and a vertex $t \in \tokenset$, define $d_t(\config)$ to be the geodesic distance of the vertices $\config^{-1}(t)$ and $\Id^{-1}(t)$ in the graph $G$. Then, $d_{max} = \max_{t\in \tokenset} d_t(\config_0)$ is a natural lower bound to the \PTS problem since for any token $t$, the distance $d_t$ can decrease by at most one using a single matching.

Lastly, we introduce some additional notation to discuss the colored version of \PTSx. Let $L = \{1,2,\ldots,\ell\}$ be a set of color labels. A labeling of the tokens is a function $\mathcal{L}: T \to L$. Then, the colored \PTS problem can be stated as, \\

\begin{problem}
    \problemtitle{\large \textit{Colored Parallel Token Swapping Problem}}
    \probleminput{A graph $G = (V,E)$, an initial configuration $\config$ and a labeling of the tokens $\mathcal{L}$.}
    \problemquestion{A sequence of matchings $(\matching_1,\ldots,\matching_k)$ with minimum size ($k$) such that for any vertex $v \in G$, $\mathcal{L}(\matching_k\cdots\matching_1\config(v)) = \mathcal{L}(\Id(v))$.}
\end{problem}
\bigskip

We note that \PTS can be cast as a simple linear algebra problem. A configuration $\config$ is just a permutation matrix on the set of vertices $V$. Each matching $M$ of the graph $G$ corresponds to a symmetric permutation matrix $P_M$ where nodes $v$ and $w$ are swapped if edge $(v,w)\in M$. Then the goal is to find a sequence of matchings of minimum size such that the matrix multiplication $M_k\cdots M_1\cdot \config$ is the identity matrix.

We conclude this section with a simple fact in the area of parallel sorting. Suppose we wish to sort $n$ numbers $a_1,\ldots,a_n$ in a list into ascending order and are allowed to swap the positions of pairs of numbers in parallel. Then the following algorithm, often called the \textit{odd-even algorithm}, sorts the numbers in at most $n$ steps \cite{knut1973art}. The algorithm alternates between odd and even phases. In an odd phase, the algorithm looks at all numbers in odd positions $a_1,a_3,\ldots$ and swaps them with the next number if $a_1>a_2$ and so on. In an even phase, the algorithm instead looks at numbers in even positions $a_2,a_4,\ldots$. Since \PTS on a line is equivalent to parallel sorting, we immediately get the following observation.

\begin{observation}\label{obs:line}
    The \PTS problem can be solved on a line graph with $n$ vertices using the odd-even algorithm (Algorithm~\ref{alg:oddeven}) in at most $n$ steps.
\end{observation}

\begin{algorithm}
    \caption{Odd-Even Algorithm}\label{alg:oddeven}
    \begin{algorithmic}[1]
        \Require A configuration $\config$ on a line graph $P_n$
        \State $E_1 \gets \{(i,i+1) : i \text{ is odd}\}$, $E_2 \gets \{(i, i+1) : i \text{ is even}\}$
        \State $\config_0 \gets \config$
        \State $k \gets 0$
        \While{$\config_k$ is not $\Id$}
            \State $k \gets k + 1$
            \If{$k$ is odd}
                \State $M_k = \{(i, i+1) \in E_1 : \config_{k-1}(i) > \config_{k-1}(i+1)\}$
            \Else
                \State $M_k = \{(i, i+1) \in E_2 : \config_{k-1}(i) > \config_{k-1}(i+1)\}$
            \EndIf
            \State $\config_k \gets M_k\config_{k-1}$
        \EndWhile
        \State \Return $M_1, \dots, M_k$
    \end{algorithmic}
\end{algorithm}

In the rest of this paper, we prove the results listed in Section~\ref{subsec:Contributions}.

\section{The stretch factor}

For a given configuration $\config$ and a vertex $t \in \tokenset$, let $d_t(\config)$ be the geodesic distance of the vertices $\config^{-1}(t)$ and $\Id^{-1}(t)$ in the graph $G$. As observed in Section~\ref{sec:prelim}, the maximum distance $d_{max}$ is a natural lower bound for \PTSx. In this section, we prove results about the stretch factor of lower bounds for \PTSx. 
Recall that the stretch factor refers to the maximum ratio of $OPT/LB$ over all instances where $OPT$ is an optimal solution and $LB$ is the lower bound. We begin by showing that any valid lower bound $f(d_1,\ldots,d_n)$ has a stretch factor of $\Omega(n)$ for general $n$-vertex graphs.

\renewcommand{\theoldthm}{\ref{thm:hardness}}
\begin{oldthm}
    Let $f(d_1,\ldots,d_n)$ be a valid lower bound for \PTS for some function $f$. Then the stretch factor of this lower bound is $\Omega(n)$ for graphs on $n$ vertices.
\end{oldthm}
\begin{proof}
    We exhibit two instances of \PTS on different graphs with the same value of the lower bound $f(d_1,\ldots,d_n)$. However the optimal solution to one of the instances is 2 and the optimal solution to the other instance is $n-1$. This proves that $f(d_1,\ldots,d_n)$ evaluates to at most two, and the ratio of $OPT/f(d_1,\ldots,d_n)$ in the second instance is $\Omega(n)$. 

    \new{We will define both graphs on the vertex set $V = \{1,2,\dots,2r\} = T$, where the only difference in the two graphs is the edge set. In both cases, we will let $\config_0$ be the initial configuration where $\config_0(i) = i-1$ for $2 \leq i \leq 2r$ and $\config_0(1) = 2r$.}

    \newcount\mycount
\begin{figure}[t!]\centering
	\begin{tikzpicture}[transform shape,scale=.5, auto]
		\node (oo) at (0,0) {};
		\foreach [count=\i] \number/\tt in {3/2,4/3,5/4,6/5,7/6,8/7,1/8,2/1}{
			\mycount=\i 		\multiply\mycount by -45		
			\node[draw,circle,inner sep=0.2cm] (v\number) at (\the\mycount:4.5cm) {\Large$v_{\number}$};	
			\node[] (t\number) at (\the\mycount:5.5cm) {\Large$t_{\tt}$};
		}
		\foreach [count=\i] \x/\y in {1/2,2/3,3/4,4/5,5/6,6/7,7/8,8/1}  \draw  (v\x) edge[bend left=15] node{} (v\y);
		
		\foreach [count=\i] \x/\y in {1/2,5/6}  \draw  (v\x) edge[bend left=15,very thick, purple] node{} (v\y);
		
		\node (a) at (-3.9,-1.8) {};\node (b) at (3.9,1.8) {};
		\draw [very thick, purple, sloped, <->]  (a) edge node{\huge diametrically opposite} (b) ;
		
		\node (c) at (6,3.2) {};\node (d) at (6,-3.2) {};
		\draw [very thick, purple, ->]  (c) edge[bend left=15,very thick, purple] node{\huge clockwise} (d) ;
	\end{tikzpicture}
	\caption{Orientation of Cycle Graphs}\label{fig:cycle_orientation}
\end{figure}

    \textbf{Cycle Graph Instance} Let $G$ be a cycle graph \new{on $V$ with edge set $E=\{(i,i+1) : 1\leq i\leq 2r-1\} \cup \{(2r,1)\}$. For ease of explanation, we will say that the vertices in $V$ are arranged in a circular manner, where vertex $i+1$ is clockwise from vertex $i$ (for $1 \leq i \leq 2r-1$) and vertex $1$ is clockwise from vertex $2r$. Then the goal is to move every token one step counter-clockwise. This configuration is depicted in Figure~\ref{fig:cycle_orientation} where $v_i$ refers to vertex $i$ and $t_i$ refers to token $i$. We use this notation in all figures throughout the paper.} We argue that any solution must use at least $2r-1$ matchings by using the following charging scheme. Initially every token has a charge of zero. If a token moves clockwise, we decrease its charge by one and if a token moves counter-clockwise, we increase its charge by one. In the final configuration, each token has moved one step counter-clockwise and so has a charge of 1 (modulo $2r$). Any matching does not change the total charge summed over all tokens as the number of tokens that move clockwise is equal to the number of tokens that move counter-clockwise. Since the total charge on all the tokens in the final configuration is zero, and every token has a charge of 1 (modulo $2r$), at least one of the tokens must have a negative charge. This charge must be $-2r+1$ or smaller. Additionally, any matching changes the charge on a token by at most $\pm 1$ and so we need at least $2r-1$ matchings. We also observe that the optimal solution uses exactly $2r-1$ matchings. 
    Set $M_i = (i,i+1)$ for $i=1,\ldots, 2r-1$ and then the sequence of matchings $M_1,\ldots,M_{2r-1}$ routes $\config_0$ to $\Id$. Here $OPT = n-1$ and $d_i = 1$ for any token $i$.

    \textbf{Complete Graph Instance} Let $G$ be the complete graph on vertex set $V$. Consider the following two matchings, $M_1 = \{(i,2r-i): i=1,\ldots,r-1\}$ and $M_2 = \{(i,2r-1-i) : i=1,\ldots,r-1\}\cup\{(2r-1,2r)\}$. Then, $M_2M_1\config_0 = \Id$. Here, $OPT = 2$ and $d_i = 1$ for any token $i$ (see Figure~\ref{fig:complete_graph}).
\end{proof}

\begin{figure}[h!]\centering
	\begin{tikzpicture}[transform shape,scale=.45, auto]
		\node[draw,circle,inner sep=0.025cm] (oo) at (0,0) {};
		\foreach [count=\i] \number/\tt in {3/2,4/3,5/4,6/5,7/6,8/7,1/8,2/1}{
			\mycount=\i 		\multiply\mycount by -45		
			\node[draw,circle,inner sep=0.2cm] (v\number) at (\the\mycount:4.5cm) {\Large$v_{\number}$};	
			\node[] (t\number) at (\the\mycount:5.5cm) {\Large$t_{\tt}$};
		}
		\foreach [count=\i] \x/\y in {1/2,2/3,3/4,4/5,5/6,6/7,7/8,8/1}  \draw  (v\x) edge[bend left=15] node{} (v\y);
		\foreach [count=\i] \x/\y in {1/7,2/6,3/5} \draw [purple,thick] (v\x) edge node{\Large$M_{1}$} (v\y);
		\node (xxx) at (0,-7) {\Huge(a) $\config_0$};
	\end{tikzpicture}
	\begin{tikzpicture}[transform shape,scale=.45, auto]
		\node[draw,circle,inner sep=0.025cm] (oo) at (0,0) {};
		\foreach [count=\i] \number/\tt in {3/4,4/3,5/2,6/1,7/8,8/7,1/6,2/5}{
			\mycount=\i 		\multiply\mycount by -45		
			\node[draw,circle,inner sep=0.2cm] (v\number) at (\the\mycount:4.5cm) {\Large$v_{\number}$};	
			\node[] (t\number) at (\the\mycount:5.5cm) {\Large$t_{\tt}$};
		}
		\foreach [count=\i] \x/\y in {1/2,2/3,3/4,4/5,5/6,6/7,7/8,8/1}  \draw  (v\x) edge[bend left=15] node{} (v\y);
		\foreach [count=\i] \x/\y/\M in {1/6/2,7/8/2} \draw [purple,thick] (v\x) edge[below,sloped] node{\Large$M_{\M}$} (v\y);
		\foreach [count=\i] \x/\y/\M in {2/5/2,3/4/3} \draw [purple,thick] (v\x) edge[above,sloped] node{\Large$M_{\M}$} (v\y);
		
		\node (xxx) at (0,-7) {\Huge(b) $M_1\config_0$};
	\end{tikzpicture}
	\begin{tikzpicture}[transform shape,scale=.45, auto]
		\node[draw,circle,inner sep=0.025cm] (oo) at (0,0) {};
		\foreach [count=\i] \number/\tt in {3/2,4/3,5/4,6/5,7/6,8/7,1/8,2/1}{
			\mycount=\i 		\multiply\mycount by -45		
			\node[draw,circle,inner sep=0.2cm] (v\number) at (\the\mycount:4.5cm) {\Large$v_{\number}$};	
			\node[] (t\number) at (\the\mycount:5.5cm) {\Large$t_{\number}$};
		}
		\foreach [count=\i] \x/\y in {1/2,2/3,3/4,4/5,5/6,6/7,7/8,8/1}  \draw  (v\x) edge[bend left=15] node{} (v\y);
		
		\node (xxx) at (0,-7) {\Huge(c) $M_2M_1\config_0 = Id$ };
	\end{tikzpicture}
	\caption{Qubit Routing on a complete graph using two matchings.}\label{fig:complete_graph}
\end{figure}

Theorem~\ref{thm:hardness} above shows that specific graph classes have to be studied to obtain stretch factors better than $O(n)$. Furthermore, the proof immediately gives Theorem~\ref{thm:stretch}(b).

\renewcommand{\theoldthm}{\ref{thm:stretch}(b)}
\begin{oldthm}
The stretch factor of $d_{max}$ for $n$-cycle graphs is between $n$ and $n-1$.
\end{oldthm}

\begin{proof}
    Consider the cycle graph instance from the proof above. Here $OPT = n-1$ while $d_{max} = 1$. Hence, the stretch factor of the lower bound $d_{max}$ is at least $n-1$. Furthermore, we can ignore an edge of the graph, say $(2r,1)$ and treat the problem as an instance on the line graph. Then, by Observation~\ref{obs:line}, any instance on the cycle graph can be solved in at most $n$ steps. Hence, the stretch factor of the lower bound $d_{max}$ is at most $n$.
\end{proof}

\subsection{Stretch Factor for Line Graphs}

In this section, \new{we} analyze the stretch factor for line graphs with respect to the maximum geodesic distance $d_{max}$, and show that this stretch factor is asymptotically tight. Let the underlying graph be a line graph $P_n = (V,E)$ where \new{$V = \{1, 2, \dots, n\}$} and $E = \{(i, i+1): 1 \leq i \leq n-1\}$. With this topology, \PTS can be viewed as a parallel version of bubble sort. We show that the stretch factor of $d_{max}$ is 2 for line graphs by proving that the odd-even algorithm achieves this bound.


We proceed in three steps. First, we define a potential function for each token. Then we show that performing an odd-even matching iteration decreases a positive potential value by at least one for each token (except perhaps on the first iteration). 
And we finally argue that the potential of any token is bounded by $2d-1$. 

We create a potential function $\Pot_\config$ for every token in the configuration $\config$. 
Informally, for any token $t$ we want to judge approximately how many iterations \new{are required to arrive at a configuration in which no tokens wish to cross $t$.}
Although the following description of the potential function is heavy on notation, the underlying ideas are actually not complex. In simple terms, for each possible prefix $[1,j]$ of vertices, we count the number of tokens in the prefix that wish to cross the token at $i$ along with the tokens on vertices in $[j+1,i-1]$ that do not wish to cross the token at $i$, and choose the largest count over all prefixes. After doing the same for all suffixes, combining the two counts yields our potential value. 

Now we will define the potential function more formally. 
Given a vertex $i$ with $\config(i) = t$, we define the set of prefixes $Q_\config^-(t)$ as the sets of \new{vertices $\{[1,j] : \config(i') > t\text{ for some }i' \in [1,j], j<i\}$} where at least one present token wishes to cross $t$. 
Similarly, we define \new{the set of} suffixes $Q_\config^+(t)$ as the sets of vertices $\{[j, n] : \config(i') < t\text{ for some }i' \in [j,n], j>i\}$. For a prefix $Q^- \in Q_\config^-(t)$ where $Q^- = [1,j]$, we let $\overline{Q^-} = [j+1, i-1]$, and similarly for a suffix $Q^+ = [j,n]$, we let $\overline{Q^+} = [i+1,j-1]$. 

To construct our potential function, we need to count tokens that must cross our token $t$, and those that do not. Let $\phi_\config^-(t, Q^-) = \left| \{i' \in Q^- : \config(i') > t\} \right|$ be the number of tokens in a prefix $Q^-$ that must cross $t$, and let $\psi_\config^-(t, \overline{Q^-}) = \left| \{i' \in \overline{Q^-} : \config(i') < t\} \right|$ be the number of tokens in $\overline{Q^-}$ that do not need to cross $t$. For suffixes, we similarly define $\phi_\config^+(t, Q^+) = \left| \{i' \in Q^+ : \config(i') < t\} \right|$ and $\psi_\config^+(t, \overline{Q^+}) = \left| \{i' \in \overline{Q^+} : \config(i') > t\} \right|$.

Now, we can write our potential function
\begin{align*}
    \Pot_\config(t) &= \max_{Q^- \in Q_\config^-(t)} \left(
\phi_\config^-(t,Q^-) + \psi_\config^-(t,\overline{Q^-}) \right) + \max_{Q^+ \in Q_\config^+(t)} \left( \phi_\config^+(t,Q^+) + \psi_\config^+(t,\overline{Q^+}) \right).
\end{align*}

For ease of notation, we will let $\Phi_\config^-(t)$ and $\Phi_\config^+(t)$ be the prefix and suffix terms of $\Phi_\config(t)$ respectively, so $\Phi_\config(t) = \Phi_\config^-(t) + \Phi_\config^+(t)$.

We now review the odd-even algorithm as presented in \cite{kawahara2017}, which finds a solution with size at most $OPT + 1$ where $OPT$ is the size of an optimal solution. The full algorithm is provided in Algorithm~\ref{alg:oddeven}. We begin by partitioning the edges in our graph into odd edges ($E_1$) and even edges ($E_2$). Then, in iteration $k$, we greedily swap tokens on odd edges if $k$ is odd, and even edges if $k$ is even, which produces the intermediate configuration $C_k$. The algorithm proceeds in this manner until it reaches the identity configuration. 


\begin{lemma}
    After an odd (resp. even) iteration of Algorithm~\ref{alg:oddeven}, all prefixes $Q^- = [1,j]$ that maximize $\phi_{\config_k}^-(t,Q^-) + \psi_{\config_k}^-(t,\overline{Q^-})$ end on even (resp. odd) $j$, and all suffixes $Q^+ = [j,n]$ that maximize $\phi_{\config_k}^+(t,Q^+) + \psi_{\config_k}^+(t,\overline{Q^+})$ begin on odd (resp. even) $j$. 
    \label{lem:oddevenmaxl}
\end{lemma} 

\begin{proof}
    Suppose we have just performed an odd iteration of Algorithm~\ref{alg:oddeven} and consider any prefix $Q^- = [1,j]$ where $j$ is odd. If $\config_k(j) < t$, then the interval $[1,j-1]$ is strictly better than $Q^-$.
    If instead $\config_k(j) > t$, then we must have $\config_k(j+1) > \config_k(j) > t$, otherwise the tokens at $j$ and $j+1$ should have swapped in iteration $k$. Therefore, the interval $[1,j+1]$ is strictly better.
    In both cases, a prefix that ends on odd $j$ cannot be the maximum.


    Symmetric arguments hold for replacing prefixes with suffixes, and following even iterations instead of odd iterations. 
\end{proof}

Next, we argue that after the first iteration of Algorithm~\ref{alg:oddeven}, the positive potential on any token decreases every iteration. 
\begin{lemma}
In running Algorithm~\ref{alg:oddeven}, if $k \geq 1$, for any token $t$ where $\Pot_{\config_k}(t) > 0$, $\Pot_{\config_k}(t) - \Pot_{\config_{k+1}}(t) \geq 1$.
\label{lem:linepotdecrease}
\end{lemma}
\begin{proof}
The proof proceeds by casework over the movement of token $t$. Take any $t$, and suppose $k$ is odd. Let us just look at the prefix side of the potential and let $Q^-$ be a prefix that maximizes $\phi_{\config_{k+1}}^-(t,Q^-) + \psi_{\config_{k+1}}^-(t,\overline{Q^-})$. 
By Lemma~\ref{lem:oddevenmaxl}, $Q^-$ must be of the form $[1,j]$ for odd $j$. We consider three cases depending on how $t$ moved during iteration $k+1$.
\begin{itemize}
    \item \emph{Case 1:} Suppose $t$ did not move. In iteration $k+1$, the edge $(j, j+1)$ was not considered, so $\phi_{\config_{k+1}}^-(t,Q^-) = \phi_{\config_{k}}^-(t,Q^-)$ and $\psi_{\config_{k+1}}^-(t,\overline{Q^-}) = \psi_{\config_{k}}^-(t,\overline{Q^-})$. But by Lemma~\ref{lem:oddevenmaxl}, $Q^-$ cannot be maximal at iteration $k$, so $\Phi_{\config_{k+1}}^-(t) < \Phi_{\config_{k}}^-(t)$, and since these values are integers, $\Phi_{\config_{k}}^-(t) - \Phi_{\config_{k+1}}^-(t) \geq 1$.
    \item \emph{Case 2:} If $t$ moved from vertex $\ell$ after iteration $k$ to vertex $\ell+1$ after iteration $k+1$, since $\config_{k+1}(\ell)$ does not contribute to $\psi_{\config_{k+1}}^-(t,\overline{Q^-})$, the same argument holds and $\Phi_{\config_{k}}^-(t) - \Phi_{\config_{k+1}}^-(t) \geq 1$.
    \item \emph{Case 3:} Finally, suppose $t$ moved from vertex $\ell$ after iteration $k$ to vertex $\ell-1$ after iteration $k+1$. Then we still have $\phi_{\config_{k+1}}^-(t,Q^-) = \phi_{\config_{k}}^-(t,Q^-)$, but $\config_{k}(\ell-1)$ contributes to $\psi_{\config_{k}}^-(t,\overline{Q^-})$, so $\psi_{\config_{k+1}}^-(t,\overline{Q^-}) = \psi_{\config_{k}}^-(t,\overline{Q^-}) + 1$. But again, by Lemma~\ref{lem:oddevenmaxl}, $Q^-$ cannot be maximal at iteration $k$, so $\Phi_{\config_{k}}^-(t) - \Phi_{\config_{k+1}}^-(t) \geq 0$.
\end{itemize}

The same results hold if $j$ is even, and mirrored results hold if we consider the suffixes instead of prefixes. So both $\Phi_{\config_{k}}^-(t)$ and $\Phi_{\config_{k}}^+(t)$ are non-increasing, and if $\Pot_{\config_k}(t) > 0$, at least one must decrease.
(Note that Case 3 implies that both $\Phi_{\config_{k}}^-(t) > 0$ and $\Phi_{\config_{k}}^+(t) > 0$, one of which must decrease.)
\end{proof}

\begin{lemma}
    Given any configuration $\config$ on a line graph $P_n$ with $d_{max} \geq 1$, the potential $\Pot_{\config}(t) \leq 2d_{max} - 1$ for every token $t \in T$.
    \label{lem:linepotential}
\end{lemma}

\begin{proof}
    Fix a token $t$ with $\config^{-1}(t) = v$. First, suppose that there are tokens on both sides of $t$ that wish to cross it. That is, there are vertices $i < v$ with $\config(i) > t$ and  $j > v$ with $\config(j) < t$. We note that $\Pot_{\config}(t)$ counts tokens, and the only tokens we may count lie between (inclusively) the smallest choice of $i$ and the largest choice of $j$. \new{Since $t$ also indicates the target vertex of token $t$,} the smallest $i$ can be is $t - d_{max} + 1$; any smaller, and there would be a token more than a distance of $d_{max}$ from its destination. Similarly, the largest $j$ can be is $t + d_{max} -1$. There are a total of $2d_{max} - 1$ tokens in this range, and we may count all but one (the token $t$ itself) as part of $\Pot_{\config}(t)$. So $\Pot_{\config}(t) \leq 2d_{max} - 2$.

    Now suppose instead that only one side of $t$ has tokens that wish to cross it. Without loss of generality, say there is some vertex $i < v$ with $\config(i) > t$. In this scenario, the tokens we count are restricted to lie between the smallest choice of $i$ and $v-1$. Once again, $i \geq t-d_{max}+1$, and clearly $v \leq t + d_{max}$. So in total, our potential function is bounded by $2d_{max} -1$.

    If no tokens wish to cross $t$, then  \new{$\Pot_{\config}(t) = 0$}, and the result is trivially true. In all cases, we have that $\Pot_{\config}(t) \leq 2d_{max} - 1$.
\end{proof}

\renewcommand{\theoldthm}{\ref{thm:stretch}(a)}
\begin{oldthm}
    The stretch factor of $d_{max}$ for line graphs is 2.
\end{oldthm}

\begin{proof}
    Consider any configuration $C$ on a line graph $P_n$ with maximum geodesic distance $d_{max}$. Lemma~\ref{lem:linepotential} says that the potential on any token is at most $2d_{max}-1$. If we perform Algorithm~\ref{alg:oddeven}, by Lemma~\ref{lem:linepotdecrease}, as long as there is positive potential on a token, it decreases by at least one in every iteration after the first. Note that once the potential on a token reaches zero, it will forever remain zero and never be part of any swap. If we assume that the potential does not increase in the first iteration, all potentials should reach zero (and so the configuration will be the identity) after $2d_{max}$ iterations.

    We only sketch the other possibility here. It is straightforward to use the same proof technique as in Lemma~\ref{lem:linepotdecrease} to show that in the first iteration, the potential on any token $t$ can increase by at most one. And if the potential increases, we must have tokens on both sides of $t$ that wish to cross, which aligns with the case in the proof of Lemma~\ref{lem:linepotential} where $\Pot_{\config}(t) \leq 2d_{max} - 2$. This yields the same $2d_{max}$ bound on the size of our solution. 

    Thus, by using Algorithm~\ref{alg:oddeven}, we can always find a solution of size no more than $2d_{max}$, and so the stretch factor of $d_{max}$ is 2.

    To show equality, we describe a family of instances where the optimal solution requires at least $2d_{max}-1$ steps. Choose an even integer $n$ and consider the configuration on $P_{2n}$ with tokens $(n+1, n+2, \dots, 2n, 1, 2, \dots, n)$ on the vertices $(1, \dots, 2n)$. Here, $d_{max} = n$. Note that Algorithm~\ref{alg:oddeven} requires at least $2n$ steps: nothing moves the first iteration, token $n+1$ cannot move for the next $n-1$ iterations and then it must move for $n$ more iterations to reach its destination. So the optimal solution is at least $2n-1$ steps implying a stretch factor that approaches $2$ as $n$ grows large.
\end{proof}

\section{Cycle graphs}

A cycle graph is a graph with vertex set $V=\{1,2,\ldots,n\}$ and edge set $E = \{(i,i+1) : 1\leq i \leq n\}$ with the understanding that $n+1 = 1$. When $n$ is even, we provide an algorithm that outputs a solution with value at most $2OPT$ and when $n$ is odd, we provide an algorithm with value at most $2OPT +1$. To design our algorithms, we introduce some definitions. Assume $n$ is even and so $n=2r$ for some integer $r$.

Let $e = (i,i+1)$ be an edge in $E$. \new{Define pairs of diametrically opposite edges to be $\left\{(i,i+1), (r+i, r+i+1)\right\}$ for $i=1,2,\ldots,r-1$, and $\left\{(r,r+1), (2r,1)\right\}$ (see Figure~\ref{fig:cycle_orientation}).} Let $P_e$ be the line graph obtained by deleting from $G$ the edge diametrically opposite to edge $e$. We say that an edge $e = (i,i+1) \in E$ is \textbf{reasonable} with respect to a configuration $\config$ if the tokens $\config(i)$ and $\config(i+1)$ on the line graph $P_e$ occur in order opposite to what their order on this path would be in the identity configuration. As an example, consider the edge $e=(r,r+1)$. The diametrically opposite edge is $(2r,1)$ and deleting this we obtain the path $(1,2,3,\ldots,2r)$. Thus, edge $e = (r,r+1)$ is \textbf{reasonable} with respect to a configuration $\config$ if and only if $\config(r) > \config(r+1)$. Consider a modification of Algorithm~\ref{alg:oddeven} that alternates between swapping all reasonable odd edges and all reasonable even edges. Call this Algorithm~\ref{alg:cycleoddeven}, specified below.

\begin{algorithm}[H]
    \caption{Cycle Odd-Even Algorithm}\label{alg:cycleoddeven}
    \begin{algorithmic}[1]
        \Require An initial configuration $\config$ on an even cycle graph $G = (V,E)$ with $n=2r$ vertices.
        \State $E_1 \gets \{(i,i+1) : i \text{ is odd}\}$, $E_2 \gets E\setminus E_1$
        \State $\config_0 \gets \config$
        \State $k \gets 0$
        \While{$\config_k\neq \Id$ and $k<n$}
            \State $k \gets k + 1$
            \If{$k$ is odd}
                \State $\matching_k = \{e \in E_1 : e \text{ is reasonable w.r.t. }\config_{k-1}\}$
            \Else
                \State $\matching_k = \{e \in E_2 : e \text{ is reasonable w.r.t. }\config_{k-1}\}$
            \EndIf
            \If{$k>1$ and $\matching_{k-1} = \emptyset$ and $\matching_{k} = \emptyset$}
                \State\Return \textit{ERROR}
            \EndIf
            \State $\config_k \gets M_k\config_{k-1}$
        \EndWhile
        \If{$\config_k\neq \Id$}
            \State\Return \textit{ERROR}
        \Else
            \State\Return $(M_1, \dots, M_k)$
        \EndIf
    \end{algorithmic}
\end{algorithm}

However, this algorithm does not always work for cycle graphs. The issue is that there may not be a feasible solution that only makes reasonable swaps. For instance, if each token wanted to move one step clockwise to arrive at its final configuration, then no swap is reasonable. We observe that this happens only in cases when $OPT$ is at least $n/2 = r$. This observation provides Algorithm~\ref{alg:evencycle} which we show to be a 2-approximation algorithm.

\begin{algorithm}
    \caption{Even Cycle Algorithm}\label{alg:evencycle}
    \begin{algorithmic}[1]
        \Require An initial configuration $\config$ on an even cycle graph $G = (V,E)$ with $n=2r$ vertices.
        \State $M_1,\ldots,M_{k_1} \gets$ Output of Algorithm~\ref{alg:cycleoddeven} on graph $G$ and initial configuration $\config$.
        \State $M'_1,\ldots,M'_{k_2} \gets$ Output of Algorithm~\ref{alg:oddeven} on graph $G' = (V,E\setminus\{(2r,1)\})$ and configuration $\config$.
        \If{ Output of 1. is \textit{ERROR} or $k_1>k_2$}
            \State \Return $M'_1,\ldots,M'_{k_2}$
        \Else
            \State \Return $M_1,\ldots,M_{k_1}$
        \EndIf
    \end{algorithmic}
\end{algorithm}


\renewcommand{\theoldthm}{\ref{thm:cycle}}
\begin{oldthm}
    Let $G$ be an $n$-cycle, and $OPT$ be the optimal solution value to an instance of \PTS on $G$. Then,
    \begin{enumerate}
        \item[(a)]  Algorithm~\ref{alg:evencycle} outputs a solution of size at most $2OPT$ for \PTS on even-cycle graphs.
        \item[(b)]  A modified version of Algorithm~\ref{alg:evencycle} outputs a solution of size at most $2OPT+1$ for \PTS on odd-cycle graphs.
    \end{enumerate}
\end{oldthm}

\begin{proof}[Proof of part (a)]
Let $OPT = (\matching_1, \matching_2,\ldots, \matching_{k^*})$ be an optimal solution. By Observation~\ref{obs:line}, we know that $k_2 \leq 2r$. Hence if $k^* \geq r$, then Algorithm~\ref{alg:evencycle} outputs a solution of size at most $2OPT$. So assume \new{that $k^* < r$}.
\begin{claim}\label{clm:cycl1}
    There exists an optimal solution $OPT = (\matching_1, \matching_2,\ldots, \matching_{k^*})$ such that $\matching_i$ only contains edges that are reasonable with respect to the configuration $\config = \matching_{i-1}\cdots\matching_{1}\config_0$.
\end{claim}

\begin{claimproof}
    Suppose not, then matching $\matching_i$ contains an edge $e=(i,i+1)$ that is not reasonable with respect to $\config$. After swapping tokens in $\config$ using the matching $\matching_i$ the token $\config(i)$ needs to cross the token $\config(i+1)$ in the path $P_e$ that is obtained by deleting the edge diametrically opposite to the edge $e = (i,i+1)$. But since $k^* < r$, the token $\config(i)$ cannot go half-way around the cycle. Hence, at a later stage, $\config(i)$ swaps its position with $\config(i+1)$ and we can delete the original swap and this new swap to still obtain an optimal solution.
\end{claimproof}

Call solutions satisfying the condition in Claim~\ref{clm:cycl1} as \textbf{reasonable solutions}. Next, define an $(E_1,E_2)$\textit{-alternating} solution to be a sequence of matchings where $\matching_{2i-1}\subseteq E_1$ and $\matching_{2i} \subseteq E_2$ for all $i$. Here, we define $E_1$ and $E_2$ as in Algorithm~\ref{alg:cycleoddeven}: $E_1 = \{(i,i+1) : i \text{ is odd}\}$, and $E_2 = E\setminus E_1$.

\begin{claim} \label{clm:cycl2}
    There exists an $(E_1,E_2)$-alternating reasonable solution of size $k^*+1$ such that no token is routed half-way around the cycle. 
\end{claim}

\begin{claimproof}
    Let $OPT = (\matching_1, \matching_2,\ldots, \matching_{k^*})$ be a reasonable solution. We can set $\matching'_1 = \matching_1 \cap E_1$, $\matching'_2 = (\matching_1 \cup \matching_2) \cap E_2$, $\matching'_3 = (\matching_2\cup \matching_3)\cap E_1$, $\ldots$, $\matching'_{k^*+1}$, \new{where $\matching'_{k^*+1} = \matching_{k^*}\cap E_1$ if $k^*$ is even and $\matching'_{k^*+1} = \matching_{k^*}\cap E_2$ if $k^*$ is odd.} \new{Recall from Remark~\ref{rem:consecutive_matchings_nonintersecting} that $M_{i} \cap M_{i+1} = \emptyset$.} Hence, the new sequence $(M'_1,\ldots,M'_{k^*+1})$ is exactly mimicking the original sequence $M_1,\ldots,M_{k^*}$ implying that the new sequence is an $(E_1,E_2)$-alternating reasonable solution of size $k^*+1$ such that no token is routed half-way around the cycle.
\end{claimproof}

\begin{claim} \label{clm:cycl3}
    If there exists an $(E_1,E_2)$-alternating reasonable solution $M'_1,\ldots,M'_{k'}$ such that no token is routed half-way around the cycle, then the output of Algorithm~\ref{alg:cycleoddeven} $M_1,\ldots,M_{k_1}$ satisfies $k_1\leq k'$.
\end{claim}

\begin{claimproof}
    Note that $M'_1 \subseteq M_1$ since $M_1$ makes all reasonable swaps. Suppose $M_1$ makes a reasonable swap $(i,i+1)$ that $M'_1$ does not make. Then in a later matching $M'_s$, the token $\config_0(i)$ must swap with the token $\config_0(i+1)$ since the token $\config_0(i)$ does not walk half-way around the cycle. But now, we can swap $\config_0(i)$ and $\config_0(i+1)$ during the first matching $M_1'$ itself (and not swap them later). Hence we can assume $M_1 = M'_1$ and repeat this argument for $M_2$ and so on.
\end{claimproof}

Claims~\ref{clm:cycl2}~and~\ref{clm:cycl3} show that if $k^*<r$, then $k_1 \leq k^*+1$. This completes the proof of the theorem.
\end{proof}

The result for odd-cycle graphs, Theorem~\ref{thm:cycle}(b), follows from a very similar analysis. A few subtle details require additional care however.


\begin{proof}[Proof of part (b)]
     Let $G = (V,E)$ be an odd-cycle graph with vertices $V=(1,2,\ldots,n=2r+1)$ and edges $E = \{(i,i+1) : 1\leq i \leq n\}$ with the understanding that $n+1 = 1$. \new{Define pairs of diametrically opposite edges to be $\left\{(i,i+1), (r+i, r+i+1)\right\}$ for $i=1,2,\ldots,r$, and $\left\{(r+1,r+2), (2r,1)\right\}$.} We can then define reasonable swaps exactly as in the case of even cycles. Additionally, we define $E_1 = {(1,2),(3,4),\ldots, (2r-1,2r)}$, $E_3 = (2r+1,1)$ and $E_2 = E\setminus(E_1\cup E_3)$. We define a $(E_1,E_3,E_2,E_3)$-alternating solution to be one where $M_{4k+1}\subseteq E_1, M_{4k+3}\subseteq E_2$ and $M_{2k}\subseteq E_3$. 
    
     If the optimal solution $M_1,\ldots,M_{k^*}$ has a token which walks half-way around the cycle, then Algorithm~\ref{alg:oddeven} on the input $G' = (V,E\setminus\{(2r+1,1)\})$ outputs a solution of size at most $2k^* + 1$. If not, we can covert $OPT$ to an $(E_1,E_3,E_2,E_3)$-alternating solution such that no token moves half-way across the cycle and of size at most $2k^*+1$ as follows. Set $M'_1 = M_1 \cap E_1$, $M'_2 = M_1 \cap E_3$, $M'_3 = (M_1 \cup M_2) \cap E_2$, $M'_4 = M_2\cap E_3$, $M'_5 = (M_2\cup M_3) \cap E_1$ and so on \new{up to $M'_{2k^*} = M_{k^*} \cap E_3$ and $M'_{2k^*+1} = M_{k^*} \cap E_1$ if $k^*$ is even and $M'_{2k^*+1} = M_{k^*} \cap E_2$ if $k^*$ is odd.}
Next, we can modify Algorithm~\ref{alg:cycleoddeven} so that in iteration $4k+1$, we make all reasonable swaps from the edge set $E_1$, in iteration $4k+3$, we make all reasonable swaps from the edge set $E_2$, and in iteration $2k$ we make all reasonable swaps from the edge set $E_3$. As in the case of even-cycles, we argue that this algorithm outputs the best $(E_1,E_3,E_2,E_3)$-alternating solution such that no token moves half-way across the cycle. This completes the proof.
\end{proof}

\section{Subdivided star graphs}

In this section, we consider the case when the input graph is a subdivided star. A subdivided star is a graph consisting of any number of paths (that we call branches) all joined at a single endpoint. 
Suppose we have a subdivided star graph with branches $1, \dots, h$, and each branch $b$ contains $n_b$ vertices and edges. We label the center vertex $v_0$, the vertices in branch $b$ as $v_{b,1}, \dots, v_{b,n_b}$, and the edges of branch $b$ as\new{ $(v_{b,1}, v_{b,2}), \dots, (v_{b,n_b-1}, v_{b,n})$} and $(v_0, v_{b,1})$. To make our notation more convenient, for every branch $b$, we let $v_{b,0} = v_0$ be an equivalent name for the center vertex, and we let the set $B_b$ be both the set of vertices in the branch as well as the tokens that belong in the branch.

 \begin{algorithm}
     \caption{Subdivided Star Algorithm}\label{alg:star}
     \begin{algorithmic}[1]
         \Require An initial configuration $\config$ on a subdivided star $G = (V,E)$ with $h$ branches with sizes $n_1, \dots, n_h$.
         \State $\config_0 \gets \config$
         \State $k \gets 0$
         \While{$C_k(v_{b,i}) \in B_b$ and $C_k(v_{b,i+1}) \notin B_b$ for some $b$ and $i$} \Comment{Phase 1}
             \State $k \gets k+1$
             \State $M_{k} = \{(v_{b,i},v_{b,i+1}): \config_{k-1}(v_{b,i}) \in B_b, \config_{k-1}(v_{b,i+1}) \notin B_b, b\in B \}$
             \State $\config_k \gets M_k\config_{k-1}$
         \EndWhile
         \While{$\config_{k}(v_{b,1}) \notin B_b$ for some $b$}        \Comment{Phase 2}
             \State $k \gets k+1$
             \State \new{$M_k = \{(v_{b,i},v_{b,i+1}): \config_{k-1}(v_{b,i}) \in B_b, \config_{k-1}(v_{b,i+1}) \notin B_b, b\in B \}$}
             \If{$\config_{k-1}(v_0) = v_0$} \Comment{Move the center token onto any unfinished branch}
                 \State $M_k \gets M_k \cup \{(v_0, v_{b,1})\}$ for some branch in $\{b:\config_{k-1}(v_{b,1}) \notin B_b\}$ 
             \Else \Comment{Swap the token on $v_0$ to its desired branch}
                 \State $M_k \gets M_k \cup \{(v_0, v_{b,1})\}$ where $b$ is the branch of $\config_{k-1}(v_0)$
                 \State $v_{c,i} \gets \config_{k-1}^{-1}(v_0)$ \Comment{Keep the center token out of the way}
                 \State $M_k \gets M_k \cup \{(v_{c,i},v_{c,i+1})\}$ \textbf{if} $\config_{k-1}(v_{c,i+1}) \notin B_c$
                 \State $M_k \gets M_k \cup \{(v_{c,i-1},v_{c,i})\}$ \textbf{if} $\config_{k-1}(v_{c,i-1}) \in B_c$
                
            \EndIf
             \State $\config_k \gets M_k\config_{k-1}$
         \EndWhile
         \State $M_{k+1}, \dots, M_{k'} \gets$ combined Algorithm~\ref{alg:oddeven} output for all branches $B_b$ and $\config_k$ \Comment{Phase 3}
         \State \Return $M_1, \dots, M_{k'}$ 
     \end{algorithmic}
 \end{algorithm}

Our algorithm proceeds in three phases. In 
\new{Phase~1}, we sort every branch $b$ so that any token that does not belong on $B_b$ is closer to $v_0$ than any token that does belong on $B_b$. (Let us call a configuration \emph{branch-sorted} if it satisfies this condition.) In Phase~2, we route all tokens to the correct branch in a greedy manner. While doing this, we make sure that the center token $v_0$ returns to its destination as infrequently as possible. Finally, in Phase~3, we use Algorithm~\ref{alg:oddeven} to sort the branches independently to reach the identity configuration.

\renewcommand{\theoldthm}{\ref{thm:star}}
\begin{oldthm}
    Algorithm~\ref{alg:star} outputs a solution of size at most $4OPT + \min\{OPT,h\} + 1$ for \PTS on subdivided stars with $h$ branches.
\end{oldthm}

\begin{proof}
    We analyze the three phases of Algorithm~\ref{alg:star} separately.

    \begin{claim}\label{clm:substar1}
        Phase~1 requires at most $OPT$ iterations.
    \end{claim}
     \begin{claimproof}
         First, we note that the sorting in Phase~1 is optimal, in the sense that this is the fewest number of matchings to arrive at a branch-sorted configuration without using the center vertex. Since we divide the tokens on each branch into only two categories, this is essentially a 0-1 sorting problem. In particular, every swap we make must be made at some point, and no edge we swap prevents another desirable swap from happening. 

         Next, we argue that the number of matchings required so that the tokens are branch-sorted is no more than $OPT$. Notice that in an optimal solution, we do eventually arrive at such a configuration, namely when all tokens are on their correct branches. 
         To properly prove the claim, we must exempt the possibility that using the center vertex could provide a significant advantage. Consider an optimal solution $M_1^*, \dots, M_{k^*}^*$, and construct a new sequence of matchings $N_1, \dots, N_{k^*}$ where $N_i$ are the edges in $M_i^*$ that are not incident to $v_0$ and when applied to $N_{i-1}\dots N_1\config_0$, they move a token on the correct branch away from the center. Then, since in any solution, any token $t_1$ on the incorrect branch must move to and through the center vertex, no token $t_2$ on the correct branch can be between $t_1$ and $v_0$. So $N_{k^*}\dots N_1\config_0$ is branch-sorted. And as we have demonstrated a sequence of matchings of size $OPT$ to arrive at a branch-sorted configuration without using the center vertex, Phase~1 of our algorithm requires no more than $OPT$ iterations.
     \end{claimproof}    
    
    \begin{claim}
        Phase~2 requires at most $OPT + \min\{OPT,h\}$ iterations.
    \end{claim}
     \begin{claimproof}
         In any solution, at most one token gets moved onto its correct branch in each matching, so the size of any optimal solution is at least the number of tokens on the wrong branch. In Phase~2, every iteration moves one token onto its correct branch unless $v_0$ is at the center. Token $v_0$ can only arrive at the center once per branch that contains a token which needs to switch branches. This number is bounded by $\min\{OPT,h\}$, and so the total number of iterations is at most $\min\{OPT,h\}$ more than the number of tokens that must switch branches. Therefore, we can bound Phase~2 by \new{$OPT + \min\{OPT,h\}$}.
     \end{claimproof}

    \begin{claim}\label{clm:substar3}
        Phase~3 requires at most $2OPT + 1$ iterations.
    \end{claim}
     \begin{claimproof}
         We once again consider an optimal solution $M_1^*, \dots, M_{k^*}^*$. 
     Let $k_1$ be the smallest number such that $M_{k_1}^*\dots M_1^*\config_0$ has all tokens on the correct branch, and
     suppose the optimal solution uses $k_2$ more matchings after step $k_1$ (i.e. $k_1+k_2 = k^*$). 
     We will split the first $k_1$ matchings of $OPT$ into a set of $2k_1$ matchings. 

     Define \new{$E_r = \{(v_{b,i-1}, v_{b,i}) : 1 \leq i \leq r, i \leq n_b\}$ to be edges} within a radius $r$ from the center vertex $v_0$. Let $M_{i,1} = M_i^* \cap E_{k_1 - i + 1}$ and $M_{i,2} = M_i^* \setminus M_{i,1}$. Then we claim that the sequence of matchings $\mathcal{M} = M_{1,1}, \dots, M_{k_1,1}, M_{1,2}, \dots, M_{k_1,2}$ applied to $\config_0$ arrives at the same configuration as $M_{k_1}^*\dots M_1^*\config_0$. 
     Note that since the radius of our edge set decreases by one for each successive matching, swaps made outside the radius (those in $M_{1,2}, \dots, M_{k_1,2}$) cannot affect the tokens being swapped in $M_{1,1}, \dots, M_{k_1,1}$. Therefore, we can defer swaps $M_{i,2}$ until after all swaps $M_{i,1}$.

     Then all tokens are on the correct branch after the first $k_1$ matchings in $\mathcal{M}$, because there are no swaps involving the center vertex in the last $k_1$ matchings. 
     Additionally, after the first $k_1$ matchings, all tokens that were on vertices in $E_{k_1}$ are still on vertices in $E_{k_1}$.

     Let $\hat{\config}$ be the configuration that Algorithm~\ref{alg:star} arrives at after Phase~2, and let $\config'$ be the configuration $M_{k_1,1}\dots M_{1,1}\config_0$. Then $\hat{\config}$ can be routed to $\config'$ in at most $k_1$ steps: for branch $b$, either $k_1 \geq n_b$, or tokens further than $k_1$ from $v_0$ are all on the correct branch. In the latter case, nothing in Phase~1 or Phase~2 of Algorithm~\ref{alg:star} nor in $M_{1,1}, \dots, M_{k_1,1}$ involved any of these tokens, so $\hat{\config}$ and $\config'$ only differ in a radius $k_1$ from $v_0$. Therefore, Algorithm~\ref{alg:oddeven} can route one to the other in at most $k_1$ steps. From $\config'$, we can arrive at the identity configuration in $k_1 + k_2$ more matchings \new{(using $M_{1,2}, \dots, M_{k_1,2}, M^*_{k_1+1}, \dots, M^*_{k_1 + k_2}$)}. Thus, there is a solution that can route from $\hat{\config}$ to $\Id$ in at most $2k_1 + k_2 \leq 2OPT$ matchings. Since we use Algorithm~\ref{alg:oddeven}, we are guaranteed to take no more than $2OPT + 1$ iterations in Phase~3.
     \end{claimproof}



    Combining the three claims above, we conclude that Algorithm~\ref{alg:star} produces a solution of size at most $4OPT + \min\{OPT,h\} + 1$.
    
\end{proof}

    

Having provided a constant factor approximation for \PTS on subdivided star graphs, we next study the stretch factor of the maximum distance lower bound $d_{max}$. We show that the stretch factor is proportional to the number of branches in the subdivided star.

\renewcommand{\theoldthm}{\ref{thm:stretch}(c)}
\begin{oldthm}
    The stretch factor of $d_{max}$ for subdivided star graphs is $\Theta(h)$ where $h$ is the maximum degree of a vertex in the graph.
    \label{thm:linestretch}
\end{oldthm}

\begin{proof}
    First, we argue that the stretch factor of $d_{max}$ is $O(h)$ by showing that Algorithm~\ref{alg:star} requires $O(hd_{max})$ iterations. As Phase~1 and Phase~3 both take place on independent paths, both are easily bounded by $d_{max}$ iterations. In Phase~2, there can be at most $d_{max}$ tokens that need to switch branches on each branch, which gives us an upper bound of $O(hd_{max})$ iterations. 

    Now, we give an example that shows the stretch factor is $\Omega(h)$.
    Consider a star graph with $n+1$ vertices (one center vertex $v_0$ and $n$ vertices incident to it). If we start in a configuration $C$ where $C(v_0) = v_0$, but all other tokens are on the wrong vertex, then $d_{max} = 2$. But any solution will require at least $n$ steps since at most one token can pass through the center on each step. 
\end{proof}

\section{Grid graphs}

We now turn our attention to grid graphs. $G = (V,E)$ is an $h\times n$ grid graph if the vertex set is $V=\{(a,b) : 1\leq a\leq h, 1\leq b \leq n\}$ and the edge set is $E = \{\bigl((a,b),(a+1,b)\bigr) : 1\leq a \leq h-1, 1\leq b\leq n\} \cup \{\bigl((a,b),(a,b+1)\bigr) : 1\leq a \leq h, 1\leq b\leq n-1\}$. We assume that $h\leq n$. We present two different algorithms for \PTS on grid graphs, each with its own merits depending on the size of $h$.

The first algorithm simply considers a spanning subgraph $G'$ of $G$ such that $G'$ is a line graph and uses Algorithm~\ref{alg:oddeven} on $G'$ to route the tokens to their correct positions. We show that the algorithm outputs a solution of size at most \new{$2OPT + 2$} when $h=2$. Such grid graphs are also called ladder graphs.

\begin{figure}[h!]\centering
	\begin{tikzpicture}[transform shape,scale=.5, auto]
		\foreach  [count=\i] \number in {1,4,5,8}{
			\node[draw,circle,inner sep=0.2cm] (v\number) at (10,-3*\i) {\Large$v_{\number}$};	
		}
		\foreach  [count=\i] \number in {2,3,6,7}{
			\node[draw,circle,inner sep=0.2cm] (v\number) at (14,-3*\i) {\Large$v_{\number}$};	
		}
		\foreach  \i/\j in {1/2,3/4,5/6,7/8}{\draw  (v\i) edge node[above]{\Large A} (v\j);}
		\foreach  \i/\j in {2/3,4/5,6/7}{\draw  (v\i) edge node[]{\Large B} (v\j);}
		\foreach  \i/\j in {1/4,3/6,5/8}{\draw  (v\i) edge node[]{\Large $F_2$} (v\j);}
		
		\node (xxx) at (17,-7) {\Large$A\cup B=F_1$};
	\end{tikzpicture}
	\caption{The partitioning of the edge set $E$ into $F_2, A$ and $B$ in ladder graphs. $F_1 = A\cup B$.}\label{fig:ladder_edge_partioning}
\end{figure}

\begin{algorithm}
    \caption{Path-Based Algorithm for Grids}\label{alg:gridpath}
    \begin{algorithmic}[1]
        \Require An initial configuration $\config$ on a $h\times n$ grid graph $G = (V,E)$.
        \State $F_1 \gets \{\bigl((a,b),(a+1,b)\bigr) : 1\leq a\leq h-1, 1\leq b\leq n\} \cup \{\bigl((h,b),(h,b+1)\bigr) : b = 1,3,5,\ldots\} \cup \{\bigl((1,b),(1,b+1)\bigr) : b=2,4,6,\ldots\}$.
        \State $M_1,\ldots,M_{k} \gets$ Output of Algorithm~\ref{alg:oddeven} on graph $G' = (V,F_1)$ and configuration $\config$.
        \State \Return $M_1,\ldots,M_{k}$
    \end{algorithmic}
\end{algorithm}

\renewcommand{\theoldthm}{\ref{thm:grid}(a)}

\begin{oldthm}
    Algorithm~\ref{alg:gridpath} outputs a solution of size \new{at most $2OPT+2$} for \PTS on $2\times n$ grid graphs. 
\end{oldthm}

 \begin{proof}
Let $M^*_1,\ldots, M^*_{k^*}$ be an optimal solution. \new{The graph $G' = (V,F_1)$} in the algorithm is formed by picking every horizontal edge and alternating between the vertical edges in the first and last columns. This creates a line graph. Let $A = \{\bigl((1,b),(2,b)\bigr) : 1\leq b\leq n\}$ be all the horizontal edges and $B = \{\bigl((2,b),(2,b+1)\bigr) : b = 1,3,5,\ldots\} \cup \{\bigl((1,b),(1,b+1)\bigr) : b=2,4,6,\ldots\}$ be all the vertical edges in $F_1$. \new{Let $E\setminus F_1 = F_2$.} We say that a solution $M_1,\ldots,M_k$ is $(A,B)$-alternating if $M_{2i+1} \subseteq A$ and $M_{2i}\subseteq B$. 
    
     We show that there exists an $(A,B)$-alternating solution with size \new{at most $2k^*+2$.} This implies that Algorithm~\ref{alg:gridpath} outputs a solution of size at most \new{$2k^*+2$} since Algorithm~\ref{alg:oddeven} outputs the best $(A,B)$-alternating solution on line graphs (This is precisely Theorem 7 in \cite{kawahara2017}).

     We begin by expanding the optimal solution into a new feasible solution as follows: $M^*_1\cap F_1, M^*_1\cap F_2, M^*_2 \cap F_2, M^*_2\cap F_1, M^*_3\cap F_1,\ldots$. Observe that the set of edges $F_2$ is a matching in itself and so any consecutive matchings $M^*_i \cap F_2$ and $M^*_{i+1}\cap F_2$ can be combined into a single matching. Hence, we obtain the following feasible solution, $M^*_1\cap F_1, (M^*_1 \cup M^*_2)\cap F_2,M^*_2\cap F_1, M^*_3\cap F_1, (M^*_3\cup M^*_4)\cap F_2,\ldots$ \new{up to $(M^*_{k^*-1} \cup M^*_{k^*}) \cap F_2, M^*_{k^*} \cap F_1$ if $k^*$ is even and $M^*_{k^*} \cap F_1, M^*_{k^*}\cap F_2$ if $k^*$ is odd. In the above transformation, we are taking two consecutive matchings $M^*_{2i-1},M^*_{2i}$ and replacing them by three matchings $M^*_{2i-1} \cap F_1, (M^*_{2i-1}\cup M^*_{2i}) \cap F_2, M^*_{2i} \cap F_1$.} We now show how to simulate a matching in $F_2$ using only the edges in $F_1$.

     \begin{claim}\label{clm:gridpath1}
         Let $M\subseteq F_2$ and $\config$ be any configuration. Then, there \new{exist} three matchings $M_1,M_2,M_3$ such that $M_3M_2M_1\config = M\config$, and $M_1\subseteq A, M_2\subseteq B$ and $M_3\subseteq A$.
     \end{claim}
     \begin{claimproof}
         Consider the case when $M=F_2$. Then setting $M_1 = A, M_2 = B$ and $M_3=A$ proves the result \new{(see Figure~\ref{fig:simulating_F2})}. Now if $M\neq F_2$, we can reduce to the previous case by splitting the ladder graph into disjoint smaller ladder graphs, where the above case applies. These can be routed in parallel. For instance, if $M = \{\bigl((1,1),(1,2)\bigr), \bigl((2,4),(2,5)\bigr)\}$. Then, the smaller ladder graphs are obtained by taking the vertices in the first two rows, and by taking the vertices in rows 4 and 5. Then, $M_1 = M_3 = \{\bigl((1,1),(2,1)\bigr),\bigl((2,1),(2,2)\bigr),\bigl((1,4),(2,4)\bigr),\bigl((1,5),(2,5)\bigr)\}$ and $M_2 = \{\bigl((2,1),(2,2)\bigr),\bigl((1,4),(1,5)\bigr)\}$.
     \end{claimproof}

     We apply Claim~\ref{clm:gridpath1} to the feasible solution from earlier to replace matchings in $F_2$ with three matchings in $F_1$. Furthermore, we split a matching $M_i\cap F_1$ into either $M_i\cap A,M_i\cap B$ or $M_i\cap B, M_i\cap A$. We use $A_i$ to denote a matching that lies in the edge set $A$ and $B_i$ to denote a matching that lies in the edge set $B$. Then, we can change the feasible solution $M^*_1\cap F_1, (M^*_1 \cup M^*_2)\cap F_2,M^*_2\cap F_1, M^*_3\cap F_1, (M^*_3\cup M^*_4)\cap F_2,\ldots$ from earlier into a feasible solution of the following form, $(B_1,A_1),(A_2,B_2,A_3),(A_4,B_3),(B_4,A_5),(A_6,B_5,A_7),\ldots$. \new{In the above transformation, we are essentially taking two consecutive original matchings $M^*_{2i-1}, M^*_{2i}$ and replacing them with $(B_{3i-2},A_{4i-3}),(A_{4i-2},B_{3i-1},A_{4i-1}),(A_{4i},B_{3i})$. But now again, since edge sets $A$ and $B$ are matchings by themselves, we can combine consecutive matchings $A_i$ and $A_{i+1}$ into a single matching. Hence, we can replace two consecutive original matchings $M^*_{2i-1}, M^*_{2i}$ with the five matchings $(B_{3i-2}, A'_{2i-1}, B_{3i-1}, A'_{2i}, B_{3i})$ where $A'_{2i-1} = A_{4i-3} \cup A_{4i-2} $ and $A'_{2i} = A_{4i-1} \cup A_{4i} $. Note that the ends of these sequences are both matchings in $B$ and so we can combine $B_{3i-2}$ and $B_{3i-3}$ into a single matching and $B_{3i}, B_{3i+1}$ into a single matching. This will give us a sequence $B_1',A_1',B_2',A_2', \cdots$ where $B'_{2i-1} = B_{3i-3} \cup B_{3i-2}$ and $B'_{2i} = B_{3i-1}$ and $B'_{2i+1} = B_{3i} \cup B_{3i+1}$. If $k^*$ is even, the tail of this sequence ends with $A_{k^*}, B_{k^*+1}$ and has a length of $2k^*+1$. On the other hand, if $k^*$ is odd, then the tail of this sequence will end in $B_{k^*+1}, A_{k^*+1}$ and has a length of $2k^*+2$.} Thus, we have exhibited an $(A,B)$-alternating solution of size at most $2k^*+2$, completing the proof of the theorem.
 \end{proof}

 \begin{figure}[h!]\centering
 \begin{subfigure}[t]{\textwidth}
 \centering
     \begin{tikzpicture}[transform shape,scale=.5, auto]
		\foreach  [count=\i]  \v/\t in {1/4,4/1,5/8,8/5}{
			\node[draw,circle,inner sep=0.2cm] (v\v) at (10,-3*\i) {\Large$v_{\v}$};	
			\node (t\t) at (9,-3*\i) {\Large$t_{\t}$};	
		}
		\foreach  [count=\i]  \v/\t in {2/2,3/6,6/3,7/7}{
			\node[draw,circle,inner sep=0.2cm] (v\v) at (14,-3*\i) {\Large$v_{\v}$};	
			\node (t\t) at (15,-3*\i) {\Large$t_{\t}$};	
		}
		\foreach  \i/\j in {1/2,3/4,5/6,7/8}{\draw  (v\i) edge  (v\j);}
		\foreach  \i/\j in {2/3,4/5,6/7}{\draw  (v\i) edge  (v\j);}
		\foreach  \i/\j in {1/4,3/6,5/8}{\draw  (v\i) edge  (v\j);}
        \foreach  \i/\j in {1/4,3/6,5/8}{\draw [purple, thick] (v\i) edge node[]{\Large $F_2$} (v\j);}
	\end{tikzpicture}
	\begin{tikzpicture}[transform shape,scale=.5, auto]
		\node (x) at (0,0) {};	\node (a) at (0,5) {};\node (b) at (4,5) {};
		\draw [very thick, purple, ->]  (a) edge[very thick, purple] node{\Large $F_2$} (b) ;
	\end{tikzpicture}
	\begin{tikzpicture}[transform shape,scale=.5, auto]
		\foreach  [count=\i]  \v/\t in {1/1,4/4,5/5,8/8}{
			\node[draw,circle,inner sep=0.2cm] (v\v) at (10,-3*\i) {\Large$v_{\v}$};	
			\node (t\t) at (9,-3*\i) {\Large$t_{\t}$};	
		}
		\foreach  [count=\i]  \v/\t in {2/2,3/3,6/6,7/7}{
			\node[draw,circle,inner sep=0.2cm] (v\v) at (14,-3*\i) {\Large$v_{\v}$};	
			\node (t\t) at (15,-3*\i) {\Large$t_{\t}$};	
			\foreach  \i/\j in {1/2,3/4,5/6,7/8}{\draw  (v\i) edge  (v\j);}
			\foreach  \i/\j in {2/3,4/5,6/7}{\draw  (v\i) edge  (v\j);}
			\foreach  \i/\j in {1/4,3/6,5/8}{\draw  (v\i) edge  (v\j);}
		}
	\end{tikzpicture}
    \caption{Routing using the matching $F_2$}
 \end{subfigure}
 
	\vspace{1em}
    
\begin{subfigure}[t]{\textwidth}\centering
	\begin{tikzpicture}[transform shape,scale=.4, auto]
		\foreach  [count=\i]  \v/\t in {1/4,4/1,5/8,8/5}{
			\node[draw,circle,inner sep=0.2cm] (v\v) at (10,-3*\i) {\Large$v_{\v}$};	
			\node (t\t) at (9,-3*\i) {\Large$t_{\t}$};	
		}
		\foreach  [count=\i]  \v/\t in {2/2,3/6,6/3,7/7}{
			\node[draw,circle,inner sep=0.2cm] (v\v) at (14,-3*\i) {\Large$v_{\v}$};	
			\node (t\t) at (15,-3*\i) {\Large$t_{\t}$};	
		}
		\foreach  \i/\j in {1/2,3/4,5/6,7/8}{\draw  (v\i) edge  (v\j);}
		\foreach  \i/\j in {2/3,4/5,6/7}{\draw  (v\i) edge  (v\j);}
		\foreach  \i/\j in {1/4,3/6,5/8}{\draw  (v\i) edge  (v\j);}
        \foreach  \i/\j in {1/2,3/4,5/6,7/8}{\draw[purple, thick] (v\i) edge node[above]{\Large A} (v\j);}
	\end{tikzpicture}
	\begin{tikzpicture}[transform shape,scale=.4, auto]
		\node (x) at (0,0) {};	\node (a) at (0,5) {};\node (b) at (2,5) {};
		\draw [very thick, purple, ->]  (a) edge[very thick, purple] node{\Large $A$} (b) ;
	\end{tikzpicture}
	\begin{tikzpicture}[transform shape,scale=.4, auto]
		\foreach  [count=\i]  \v/\t in {1/2,4/6,5/3,8/7}{
			\node[draw,circle,inner sep=0.2cm] (v\v) at (10,-3*\i) {\Large$v_{\v}$};	
			\node (t\t) at (9,-3*\i) {\Large$t_{\t}$};	
		}
		\foreach  [count=\i]  \v/\t in {2/4,3/1,6/8,7/5}{
			\node[draw,circle,inner sep=0.2cm] (v\v) at (14,-3*\i) {\Large$v_{\v}$};	
			\node (t\t) at (15,-3*\i) {\Large$t_{\t}$};	
			\foreach  \i/\j in {1/2,3/4,5/6,7/8}{\draw  (v\i) edge  (v\j);}
			\foreach  \i/\j in {2/3,4/5,6/7}{\draw  (v\i) edge  (v\j);}
			\foreach  \i/\j in {1/4,3/6,5/8}{\draw  (v\i) edge  (v\j);}
		}
        \foreach  \i/\j in {2/3,4/5,6/7}{\draw[purple, thick]  (v\i) edge node[]{\Large B} (v\j);}
	\end{tikzpicture}
	\begin{tikzpicture}[transform shape,scale=.4, auto]
		\node (x) at (0,0) {};	\node (a) at (0,5) {};\node (b) at (2,5) {};
		\draw [very thick, purple, ->]  (a) edge[very thick, purple] node{\Large $B$} (b) ;
	\end{tikzpicture}
	\begin{tikzpicture}[transform shape,scale=.4, auto]
		\foreach  [count=\i]  \v/\t in {1/2,4/3,5/6,8/7}{
			\node[draw,circle,inner sep=0.2cm] (v\v) at (10,-3*\i) {\Large$v_{\v}$};	
			\node (t\t) at (9,-3*\i) {\Large$t_{\t}$};	
		}
		\foreach  [count=\i]  \v/\t in {2/1,3/4,6/5,7/8}{
			\node[draw,circle,inner sep=0.2cm] (v\v) at (14,-3*\i) {\Large$v_{\v}$};	
			\node (t\t) at (15,-3*\i) {\Large$t_{\t}$};	
			\foreach  \i/\j in {1/2,3/4,5/6,7/8}{\draw  (v\i) edge  (v\j);}
			\foreach  \i/\j in {2/3,4/5,6/7}{\draw  (v\i) edge  (v\j);}
			\foreach  \i/\j in {1/4,3/6,5/8}{\draw  (v\i) edge  (v\j);}
		}
        \foreach  \i/\j in {1/2,3/4,5/6,7/8}{\draw[purple, thick]  (v\i) edge node[above]{\Large A} (v\j);}
	\end{tikzpicture}
	\begin{tikzpicture}[transform shape,scale=.4, auto]
		\node (x) at (0,0) {};	\node (a) at (0,5) {};\node (b) at (2,5) {};
		\draw [very thick, purple, ->]  (a) edge[very thick, purple] node{\Large $A$} (b) ;
	\end{tikzpicture}
	\begin{tikzpicture}[transform shape,scale=.4, auto]
		\foreach  [count=\i]  \v/\t in {1/1,4/4,5/5,8/8}{
			\node[draw,circle,inner sep=0.2cm] (v\v) at (10,-3*\i) {\Large$v_{\v}$};	
			\node (t\t) at (9,-3*\i) {\Large$t_{\t}$};	
		}
		\foreach  [count=\i]  \v/\t in {2/2,3/3,6/6,7/7}{
			\node[draw,circle,inner sep=0.2cm] (v\v) at (14,-3*\i) {\Large$v_{\v}$};	
			\node (t\t) at (15,-3*\i) {\Large$t_{\t}$};	
			\foreach  \i/\j in {1/2,3/4,5/6,7/8}{\draw  (v\i) edge  (v\j);}
			\foreach  \i/\j in {2/3,4/5,6/7}{\draw  (v\i) edge  (v\j);}
			\foreach  \i/\j in {1/4,3/6,5/8}{\draw  (v\i) edge  (v\j);}
		}
	\end{tikzpicture}
    \caption{Routing using the three matchings $A, B$ and $A$}
    \end{subfigure}
	\caption{Simulating the routing from $F_2$ using the three matchings $A, B$ and $A$.}\label{fig:simulating_F2}
\end{figure}

Algorithm~\ref{alg:gridpath} works well on ladder graphs but the approximation factor increases multiplicatively with the size of $h$. Instead, we provide another algorithm whose approximation factor increases only additively with the size of $h$. The idea here is to first route the tokens so that each token lies on the same row as its final position, and then to sort the rows in parallel. Say that a token $t$ \textit{belongs} to row $i$ if the vertex $\Id^{-1}(t)$ lies on row $i$. We say that a configuration $\config$ on the grid graph is \textit{amicable} if for every column $j$ and row $i$, there exists exactly one token $t$ such that the vertex $\config^{-1}(t)$ lies on column $j$ and token $t$ belongs to row $i$. The result in \cite{alon1994} shows that we can route the initial configuration $\config_0$ into an amicable configuration $\config$ by only moving tokens along rows. Starting from this amicable configuration $\config_1$, we can move tokens along columns to reach a new configuration $\config_2$ where every token already lies in the row it belongs to. Finally, we again, move tokens along rows to obtain the final configuration $\Id$. For a configuration $\config$, define $\mathcal{A}(C)$ to be the amicable configuration obtained by moving tokens only along rows as in \cite{alon1994}. For an amicable configuration $\config$, define $\mathcal{B}(C)$ to be the configuration obtained by moving tokens only along columns so that each token lies on the row it belongs to.

 \begin{algorithm}
     \caption{3-Phase Algorithm for Grids}\label{alg:gridphases}
     \begin{algorithmic}[1]
         \Require An initial configuration $\config$ on a $h\times n$ grid graph $G = (V,E)$.
         \State $\config_1 \gets \mathcal{A}(\config)$
         \State $\config_2 \gets \mathcal{B}(\config_1)$
         \State $M_1,\ldots,M_{k_1} \gets$ The sequence of matchings that routes the rows from $\config$ to $\config_1$ using Algorithm~\ref{alg:oddeven} in parallel.
         \State $M_{k_1+1},\ldots,M_{k_2} \gets$ The sequence of matchings that routes the columns from $\config_1$ to $\config_2$ using Algorithm~\ref{alg:oddeven} in parallel.
         \State $M_{k_2+1},\ldots,M_{k} \gets$ The sequence of matchings that routes the rows from $\config_2$ to $\Id$ using Algorithm~\ref{alg:oddeven} in parallel.
         \State \Return $M_1,\ldots,M_k$
     \end{algorithmic}
 \end{algorithm}

\renewcommand{\theoldthm}{\ref{thm:grid}(b)}
\begin{oldthm}
    Algorithm~\ref{alg:gridphases} outputs a solution of size at most $2OPT+2h$ for \PTS on $h\times n$ grid graphs. 
\end{oldthm}

\begin{proof}
    Algorithm~\ref{alg:gridphases} works in three phases. We bound the number of matchings in each of the phases separately. In the first phase, the algorithm routes the configuration $\config$ into the configuration $\config_1$ using Algorithm~\ref{alg:oddeven} on the rows. Due to Observation~\ref{obs:line}, this takes at most $h$ matchings. The second phase routes the configuration $\config_1$ into the configuration $\config_2$ using Algorithm~\ref{alg:oddeven} on the columns. Let $d_{max}$ be the maximum geodesic distance between $\config^{-1}(t)$ and $\Id^{-1}(t)$ in the graph $G$ for any token $t$. Let $d'$ be the maximum geodesic distance between $\config_1^{-1}(t)$ and $\config_2^{-1}(t)$ in the graph $G$ for any token $t$. The geodesic distance between $\config_1^{-1}(t)$ and $\config_2^{-1}(t)$ for any token $t$ corresponds to the vertical distance between $\config^{-1}(t)$ and $\Id^{-1}(t)$. Hence, $d'\leq d_{max}$.
    
    Then, by Theorem~\ref{thm:stretch}(a), the second phase takes at most $2d'\leq 2d_{max} \leq 2OPT$ matchings. Finally the third phase is bounded the same way as the first and requires at most $h$ matchings. Putting the three phases together, we obtain that Algorithm~\ref{alg:gridphases} outputs a solution with at most $2OPT + 2h$ matchings.
\end{proof}

\begin{remark}
    The above proof shows that Algorithm~\ref{alg:gridphases} outputs a solution of size at most $2d_{max} + 2h$ and so the stretch factor of the lower bound $d_{max}$ on grid graphs can be bounded by $O(h)$. 
\end{remark}

\section{Colored parallel token swapping}

We now turn our attention to colored versions of \PTSx. If the number of tokens in the \PTS problem is smaller than the number of nodes, all empty positions can be assigned a same color and dummy tokens with that particular color can be added in (the dummy tokens are interchangeable by virtue of having the same color). This is called the incomplete \PTS problem. A further generalization is the colored \PTS problem where each token and each vertex is assigned a color, and every token should be routed to a vertex that shares its color.

In general, all our results for the colored parallel token swapping problem arise from reductions to \PTSx. We will construct a final feasible configuration $C^*$ with $\mathcal{L}(C^*(v)) = \mathcal{L}(Id(v))$ for all vertices $v\in G$, and use our algorithms for \PTS to route from the initial configuration $C$ to the final feasible configuration $C^*$. Observe that there are potentially many feasible final configurations $C^*$ but a swap between two identically colored tokens is redundant in an optimal sequence of matchings. Thus, we shall not consider any matching that swaps identically colored tokens and can restrict our attention to final configurations $C^*$ that do not require such a swap. This is a useful observation to keep in mind during this section. For instance, it immediately implies that there is exactly one candidate final configuration $C^*$ for the colored parallel token swapping problem on line graphs. Thus, using the result of \cite{kawahara2017}, one can obtain a solution with at most $OPT + 1$ matchings for the colored \PTS problem on line graphs. We prove the following theorem about the colored and incomplete \PTS problems on cycle, sub-divided star and grid graphs.



\renewcommand{\theoldthm}{\ref{thm:color}}
\begin{oldthm}
    \begin{enumerate}
        \item[(a)] If $G$ is a cycle, we can find a solution with value at most $2 OPT + 1$ for the incomplete \PTS in polynomial time.
        \item[(b)] If $G$ is a sub-divided star, we can find a solution with value at most $4 OPT + \min\{OPT,h\} + 1$ for the colored \PTS in polynomial time, and by extension for the incomplete \PTS problem.
        \item[(c)] If $G$ is an $h \times n$ grid graph with $h \leq n$, we can find a solution with value at most $2 OPT + 2h$ for the colored \PTS in polynomial time, and by extension for the incomplete \PTS problem.
    \end{enumerate}
\end{oldthm}

\begin{proof}[Proof of part (a)]
    Recall that in the incomplete \PTS problem, there is exactly one color label (say $\ell$) with multiple tokens belonging to that color and every other color label has exactly one token. A simple observation is that it is never beneficial to swap the positions of two tokens with the same color label. Let $\config^*$ be the final configuration after an optimal sequence of matchings where tokens of the same color have never been swapped. The positions of tokens $t$ whose color label is not $\ell$ are fixed in the final configuration to be $\Id^{-1}(t)$. For a particular token $t$ whose color label is $\ell$, there are at most $n$ possible positions for this token in the final configuration. Since tokens of the same color do not swap, fixing the position of token $t$ immediately assigns a position around the cycle to every other token with color label $\ell$ as well. Thus, we obtain a list of at most $n$ configurations such that $\config^*$ is guaranteed to be in the list. For each configuration $\config'$ in this list, we find a sequence of matchings that routes the initial configuration $\config$ to the configuration $\config'$ using Algorithm~\ref{alg:evencycle}. We output the best solution found among these. Since $\config^*$ is in the list, Theorem~\ref{thm:cycle} tells us that we have found a solution of size at most $2OPT+1$.
\end{proof}

\begin{proof}[Proof of part (b)]
    Suppose we have a subdivided star graph with branches $1, \dots, h$, and each branch $b$ contains $n_b$ vertices and edges. 
    We next construct a feasible final configuration $\config_1$ as follows:
    We will handle one color at a time. Fix a color $\ell$. For every branch $b$, let $r_b$ be the number of $\ell$-colored tokens (under $\config_0$) and $q_b$ be the number of $\ell$-colored vertices on $b$. For each branch, beginning from the end of the branch, let $v_{b,1}, v_{b,2}, \dots, v_{b,q_b}$ be the $\ell$-colored vertices in order, and let $t_{b,1}, t_{b,2}, \dots, t_{b,q_b}$ be the $\ell$-colored tokens in order. We set $\config_1(v_{b,i}) = t_{b,i}$ for $i \in \{1, \dots, \min\{r_b, q_b\}\}$ on each branch. We assign all remaining $\ell$-colored tokens arbitrarily to the remaining $\ell$-colored vertices.

    We  use a  slightly modified version of Algorithm~\ref{alg:star} to route the tokens from $\config_0$ to a correctly-colored configuration. To this end we first  run Phase~1 and Phase~2 of Algorithm~\ref{alg:star} for routing tokens from $\config_0$ to the target configuration $\config_1$. Then in Phase~3, we change our target configuration  $\config_1$ so that  tokens of the same color do not swap. 
    At the end of Phase~2, when each branch should contain the correct number of tokens of each color, we  run Algorithm~\ref{alg:oddeven} on each branch to move to the closest correctly-colored configuration (so that no tokens of the same color need to swap), which we can find greedily.
    Using this modified Algorithm~\ref{alg:star} to route the tokens provides a solution of size at most $4OPT + \min\{OPT,h\} + 1$. 
    
    The rest of the proof follows the same outline as the proof of Theorem~\ref{thm:star} where we bound the number of iterations in each phase against an optimal solution. To avoid repetition, we present the rest of the proof in the Appendix.
    

\end{proof}
  
\begin{proof}[Proof of part (c)]
    For a feasible final configuration $\config_1$ and an initial configuration $\config_0$, define the distance between these configurations $d(\config_1,\config_0)$ to be the maximum geodesic distance between $\config_1^{-1}(t)$ and $\config_0^{-1}(t)$ over all tokens $t$. Let $d^*$ be the minimum value of $d(\config_1,\config_0)$ over all feasible final configurations $\config_1$. Note that $OPT \geq d^*$ for the colored \PTS problem. We find a feasible configuration $\config_1$ that achieves this distance $d^*$ as follows.

    For $d = 1,\ldots, h+n$, and for each color label $\ell_i$, construct a bipartite graph $G'$ with vertex set $A\cup B$ where $A = \{v\in G: v = \config_0^{-1}(t) \text{ for some $t$ with color $\ell_i$}\}$ and $B = \{v\in G: v = \Id^{-1}(t) \text{ for some $t$ with color $\ell_i$}\}$. Add an edge between a vertex in $A$ and vertex in $B$ if the geodesic distance between these vertices in the graph $G$ is at most $d$. If $G'$ contains a perfect matching, then it implies that there exists a configuration $\config_1$ so that the distance between tokens of color $\ell_i$ between $\config_1$ and $\config_0$ is at most $d$. Thus, we can find the smallest $d$ so that this holds true for every color label.

    Having found the desired configuration $\config_1$ with minimum $d(\config_1,\config_0)$, we apply Algorithm~\ref{alg:gridphases} to route $\config_0$ to $\config_1$. As analyzed in Section 6, the number of matchings required is at most $2d^*+2h$ and since $OPT\geq d^*$, we obtain our result.
\end{proof}

\begin{remark}
    Our methodology for incomplete \PTS on cycle graphs do not extend to the colored \PTS on cycle graphs. This is because each color could have $\Omega(n)$ possible final configurations leading to $\Omega(n^{|\mathcal{L}|})$ possible final configurations for all the tokens. We cannot enumerate these as we do for incomplete \PTS and a new approach needs to be developed. This is an open question.
\end{remark}

\section{Concluding remarks and future work}

The parallel token swapping is more complex than the regular token swapping problem and approximation algorithms might need to be developed keeping the specific network topology in mind. Theorem~\ref{thm:hardness} provides some evidence for this difficulty arguing that natural distance based lower bounds are insufficient for designing approximation algorithms since their stretch factor can be arbitrarily bad. Nonetheless, we show that it is indeed possible to use properties of the underlying network topology to design good approximation algorithms and do so for cycle graphs, subdivided star graphs and grid graphs. We hope that our work along these lines will motivate future research into the parallel token swapping problem. For instance, do our techniques for subdivided star graphs extend to general trees? Or do our techniques for cycle graphs extend to outerplanar graphs?

In addition to designing approximation algorithms, we study the stretch factor of the maximum distance lower bound and provide tight bounds for the same in the context of line graphs, cycle graphs and subdivided star graphs. While we provide an upper bound for the stretch factor on grid graphs, pinning it down asymptotically remains an open question. The study of the stretch factor is motivated from operations research applications where one can use lower bounds as a proxy for the true optimal parallel token swapping sequence. The advantage is that the maximum distance lower bound can be expressed as a linear term allowing for simple mixed integer linear programming formulations. This idea was utilized, for instance, in the qubit routing problem where the distance sum lower bound for the regular token swapping problem was utilized to obtain well-performing heuristics to minimize size in the more complex qubit routing problem \cite{wagner2023improving}. We hope that our work on the parallel token swapping problem will lead to heuristics that minimize the depth of the qubit routing problem as alluded to in Section~\ref{sec:motivation}.

\bibliographystyle{plain}
\bibliography{arxiv}

\section*{Appendix: Proof of Theorem~\ref{thm:color}(b)}

   \begin{claim}\label{clm:substarcolor0}
        For every branch $b$ and every color $\ell$, the number of $\ell$-colored tokens on branch $b$ in $\config_0$ and not on $b$ in $\config_1$ is the minimum possible over all feasible final configurations.
    \end{claim}
    \begin{claimproof}
        Fix color $\ell$. Consider branch $b$ with $r$ $\ell$-colored tokens and $q$ $\ell$-colored vertices. If $r \geq q$, then the minimum number of $\ell$-colored tokens that must leave the branch is $r-q$. This is exactly the number that leave under $\config_1$. Similarly, if $q \geq r$, the minimum number of $\ell$-colored tokens that must move to the branch is $q-$. Again, this is exactly the number that move to this branch under $\config_1$.
    \end{claimproof}

  \begin{claim}\label{clm:substarcolor2}
        Phase 1 requires at most OPT iterations.
    \end{claim}
    \begin{claimproof}
        As in the proof of Claim~\ref{clm:substar1}, we know that Phase 1 is the optimal sequence of matchings (of length $k$) to arrive at a branch-sorted configuration with respect to $\config_1$ without using the center vertex. 
        But we would like to argue that there is no better sequence to arrive at a branch-sorted configuration with respect to any other feasible final configuration $\config'$. 

        Given a configurations $\config$ and $\config'$, let $T(\config,\config')$ be the set of tokens that are on different branches in $\config$ and $\config'$. Now suppose there is some other sequence of matchings $M'_1, \dots, M'_{k'}$ not using the center vertex with $k' < k$ that arrives at a branch-sorted configuration with respect to some minimal $\config' \neq \config_1$, where minimality is in terms of the following function:
        $$\sum_{t \in T(\config_0, \config')} d(\config_0^{-1}(t), v_0).$$
        Then for some branch $b$, the set of tokens $B^-_b$ on $b$ in $\config_0$ that should move to a different branch in $\config_1$ is different from the set $B'^-_b$ of those that must move to a different branch in $\config'$.
        In particular, there must be some color $\ell$ such that the $\ell$-colored tokens in $B^-_b$ are different from the $\ell$-colored tokens in $B'^-_b$.

        By Claim~\ref{clm:substarcolor0}, if $B^-_b$ has $r$ $\ell$-colored tokens, $B'^-_b$ must have at least $r' \geq r$. If the number is strictly greater, let $t$ be the $\ell$-colored token in $B'^-_b$ farthest from $v_0$. It is straight-forward to find a new configuration $\config''$ where $t$ is not in $B'^-_b$, and we can arrive a configuration branch-sorted with respect to $\config''$ by ignoring all swaps that included $t$. (If $t$ is the $q$th $\ell$-colored token from the end of the branch, such a configuration can be found by setting $\config''(v_q) = t$, where $v_1$ is the $q$th $\ell$-colored vertex, and $\config''(\config'^{-1}(t)) = \config'(v_q)$.) This contradicts the minimality of $\config'$.
        
        If instead, $r' = r$, then there must be $\ell$-colored tokens $t_1$ and $t_2$ where $t_1 \in B^-_b$ and $t_2 \in B'^-_b$, but $t_1 \notin B'^-_b$ and $t_2 \notin B^-_b$. By the way we constructed $\config_1$, $t_1$ must be farther from $v_0$ than $t_2$. So $t_1$ and $t_2$ must have swapped in $M'_1, \dots, M'_{k'}$. So if we define a new configuration $\config''$ as the same except $\config''(\config'^{-1}(t_2)) = t_1$ and $\config''(\config'^{-1}(t_1)) = t_2$, then we can arrive at a configuration branch-sorted with respect to $\config''$ by performing the same matching sequence except for the swap of $t_1$ and $t_2$. This also contradicts the minimality of $\config'$.
    
        Therefore, Phase 1 is an optimal matching sequence excluding the center vertex to find a branch-sorted configuration with respect to any feasible final configuration.

        Finally, we repeat the argument from Claim~\ref{clm:substar1} to construct a sequence of matchings of size $OPT$ without using the center vertex to arrive at some branch-sorted configuration. And so, Phase 1 requires no more than $OPT$ iterations.
    \end{claimproof}

    \begin{claim}\label{clm:substarcolor3}
        Phase 2 requires at most $OPT + \min\{OPT,h\}$ iterations.
    \end{claim}
    \begin{claimproof}
        In every iteration of Phase 2, one token moves to a new branch (and never switches branches again), unless the token at $v_0$ is $C_1(v_0)$, which could happen up to $\min\{OPT,h\}$ times. By Claim~\ref{clm:substarcolor0}, $\config_1$ is the configuration that requires the minimum number of tokens to move to a new branch from $\config_0$. So any solution must require at least that many matchings.
        In Phase 2, we may use up to $\min\{OPT,h\}$ more matchings than this, so Phase 2 requires at most $OPT + \min\{OPT,h\}$ iterations. 
    \end{claimproof}

    \begin{claim}\label{clm:substarcolor4}
        Phase 3 requires at most $2OPT + 1$ iterations.
    \end{claim}
    \begin{claimproof}
    Consider an optimal solution  $M_1^*, \dots, M_{k^*}^*$ and let $\config^* = M_{k^*}^*\dots M_1^*\config_0$.
    Define $k_1$ as the smallest number such that no matching in $M_{k_1}^*, \dots M_{k^*}^*$ includes an edge adjacent to $v_0$. And suppose the optimal solution uses $k_2$ more matchings after step $k_1$ (i.e. $k_1+k_2 = k^*$). 
    As in the proof of Claim~\ref{clm:substar3}, we will split the first $k_1$ matchings of $OPT$ into a set of $2k_1$ matchings. 

    Define $E_r = \{(v_{b,i-1}, v_{b,i} : 1 \leq i \leq r\}$ be edges within a radius $r$ from the center vertex $v_0$. Let $M_{i,1} = M_i^* \cap E_{k_1 - i + 1}$ and $M_{i,2} = M_i^* \setminus M_{i,1}$. Then we can argue as in the non-colored version that the sequence of matchings $\mathcal{M} = M_{1,1}, \dots, M_{k_1,1}, M_{1,2}, \dots, M_{k_1,2}$ applied to $\config_0$ arrives at the same configuration as $M_{k_1}^*\dots M_1^*\config_0$.
     

    Let $\hat{\config}$ be the configuration that Algorithm~\ref{alg:star} arrives at after Phase~2, and let $\config'$ be the configuration $M_{k_1,1}\dots M_{1,1}\config_0$. Note that in both cases, on every branch, the number of tokens of each color matches the number of vertices of each color. We claim that any tokens more than a distance $k_1$ from the $v_0$ did not move in Phase~1 and Phase~2. 

    Suppose that some token more than distance $k_1$ from $v_0$ did move in Phase~1 or Phase~2. This means there is some token $t$ on branch $b$ that is at least $k_1$ units from $v_0$ in $\config_0$ and is on a different branch in $\config_1$. If $t$ is of color $\ell$, by the construction of $\config_1$, all other tokens on branch $b$ in $\config_0$ that are closer to $v_0$ than $t$ must also be on a different branch in $\config_1$. Furthermore, if in $\config_0$, $t$ is the $r$th $\ell$-colored token on branch $b$ from $v_0$, at least $r$ such tokens must move to a different branch in any valid final configuration. In particular, either token $t$ or an $\ell$-colored token even farther from $v_0$ must move to a different branch. But we know there is a solution where $t$ does not move to another branch: $\config'$. This is a contradiction, so we conclude that tokens more than a distance $k_1$ from $v_0$ did not move in Phase~1 or Phase~2.
    
    So $\config'$ and $\hat{\config}$ only differ at vertices within a distance $k_1$ from $v_0$. Therefore we can route $\hat{\config}$ to a configuration equivalent (in terms of colors) to $\config'$ in at most $k_1$ steps, using Algorithm~\ref{alg:oddeven} on each branch separately.
    Finally, we can arrive at the identity configuration in $k_1 + k_2$ more matchings, using $M_{1,2}, \dots, M_{k_1,2}, M_{k_1+1}, \dots, M_{k_1 + k_2}$. So there is a solution that can route from $\hat{\config}$ to $\Id$ in at most $2k_1 + k_2 \leq 2OPT$ matchings. And the number of matchings needed to route each branch to a correctly-colored solution that requires no same-color token swapping is at least as good. Since we use Algorithm~\ref{alg:oddeven}, we require at most $2OPT+1$ iterations in Phase~3.
    \end{claimproof}

    Combining the claims above, we conclude that Algorithm~\ref{alg:star} produces a solution of size at most $4OPT + \min\{OPT,h\} + 1$.

\end{document}